\documentclass[%
reprint, 
 amsmath,amssymb,
 aps,
]{revtex4-2}

\usepackage{alan_preamble}
\usepackage{soul}
\title{Main file}
\date{August 2022}

\begin{document}

\preprint{APS/123-QED}

\title{Field-level multiprobe analysis of the CMB, integrated Sachs-Wolfe effect, and the galaxy density maps}

\author{Alan Junzhe Zhou}
\author{Scott Dodelson}%
\affiliation{%
 Department of Physics, Carnegie Mellon University, Pittsburgh, PA 15213.\\
 McWilliams Center for Cosmology, Carnegie Mellon University, Pittsburgh, PA 15213.\\
 NSF AI Planning Institute, Carnegie Mellon University, Pittsburgh, PA 15213.
}%

\date{\today}

\begin{abstract}
Extracting information from cosmic surveys is often done in a two-step process, construction of maps and then summary statistics such as two-point functions. We use simulations to demonstrate the advantages of a general Bayesian framework that consistently combines different cosmological experiments on the field level, and reconstructs both the maps and cosmological parameters. We apply our method to jointly reconstruct the primordial CMB, the integrated Sachs-Wolfe effect, and six tomographic galaxy density maps on the full sky on large scales along with several cosmological parameters. While the traditional maximum a posterior estimator has both two-point level and field-level bias, the new approach yields unbiased cosmological constraints and improves the signal-to-noise ratio of the maps.
\end{abstract}

\maketitle

\section{Introduction}
The large-scale structure (LSS) of the Universe is defined by the full 3-dimensional matter density field $\delta(\bfx,t)$. Although it is difficult to determine $\delta(\bfx,t)$ directly, we extract information about it indirectly in two general ways: (i) light from distant sources (including the cosmic microwave background) is impacted by over and underdense regions; and (ii) gravitationally bound objects such as galaxies and clusters often {\it trace} the matter density. Examples of the first class of information include the late-time integrated Sachs-Wolfe (ISW) effect caused by decaying gravitational potentials in the dark-energy era and the deflection of photons due to gravitational lensing. The second class includes galaxy clustering and cluster counts. One important objective of modern cosmology is to develop statistical methods to combine this information in the most efficient and consistent manner, in order to reconstruct $\delta(\bfx,t)$ and constrain models of its origin and evolution. 

In the past decades, independent experiments have made extraordinary advances in charting these individual tracers. For example, on the cosmic microwave background (CMB) front, several generations of anisotropy and polarization measurements have led to recent results; the Planck Collaboration has mapped the temperature and polarization anisotropy of the early Universe and used its lensing statistics to study the integrated gravitational potential along the line of sight~\cite{Planck2018summary,Planck2018lensing}. The Atacama Cosmology Telescope (ACT) and the South Pole Telescope (SPT) have made similar achievements with smaller footprints but higher resolutions~\cite{Aiola_2020, Darwish_2020, Bianchini_2020,https://doi.org/10.48550/arxiv.2212.05642}. Stage-III wide-field photometric surveys such as the Dark Energy Survey (DES), the KioDegree Survey (KiDS) and the Hyper Suprime-Cam (HSC) have observed millions of galaxies on a significant fraction of the sky and used galaxy positions and shape statistics to probe the low-redshift matter distributions~\cite{DESY3Cosmology,Heymans_2021,Hamana_2020}. The recipe for analyzing most of this data involves first converting the data into 2-dimensional maps (e.g., for CMB surveys) and catalogs (e.g., for galaxy surveys); computing the correlation functions (or the power spectra) of these fields, and then comparing these observed correlation statistics to a cosmological model in a Bayesian likelihood analysis to yield cosmological parameter constraints. In almost all of these cases, the fiducial cosmological model, $\Lambda$CDM, fits the data well.

In addition to these results from single probes, there has been an increased effort to maximize information by combining probes. An example of this is the recent DES result combining its data of galaxy positions and galaxy shapes with the projected gravitational potential measured by SPT and Planck~\cite{https://doi.org/10.48550/arxiv.2206.10824}.  In this example, roughly the same recipe is followed: DES made maps of the galaxy density in five tomographic bins and the shear in four bins; these were combined with maps of the projected gravitational potential from SPT and Planck. Given these three sets of maps, there are six sets of two-point functions (galaxy clustering, galaxy-galaxy lensing; cosmic shear, cosmic shear $\times$ CMB lensing, galaxy density $\times$ CMB lensing, and the CMB lensing auto-correlation function). This set of six two-point functions forms the data vector, which is then used to constrain parameters. The main goal of this effort is to extract from all this low-redshift (much lower than the decoupling of the CMB) data a measurement of the amount of clustering at late times. This is often quantified with $S_8$, which the DES+SPT analysis determined to be $S_8=0.792\pm 0.012$,  lower than the Planck measurement, $S_8=0.832\pm 0.013$. The discrepancy does not meet strict statistical standards but it has spawned much interest and it is reminiscent of the Hubble tension that is driven by different measurements of the zeroth order expansion rate of the universe. 

Taking stock, the fiducial cosmological model fits most of the data, but there are alluring hints that it is flawed, and one of the most intriguing ways of stress-testing the model is to measure how the clustering of matter evolves over the course of time. To date, this has been done predominantly by: (i) map-making, (ii) compression to two-point functions, and (iii) parameter constraints. 

Research into field-level analysis offers an opportunity to change the way that we extract data from surveys, in the process offering an alluring opportunity for a powerful suite of tests of $\Lambda$CDM.  The basic idea of field-level analysis is to combine all three steps above into one. Early examples of this idea~\cite{Tegmark_1997,wandelt_globalexactCMBGibbs2003,Larson_2007,eriksen_BayesianCompSepCl2008,larsonEstimationPolarizedPower2007a,anderesBayesianInferenceCMB2015} focused on the CMB. In that example, the time-ordered data can be converted into a map at the same time that the power spectrum is determined. The parameters to fit for the data, therefore, are the values of the temperature in all the pixels in the map plus a handful of cosmological parameters that determine the power spectrum. \citet{eriksen_BayesianCompSepCl2008} extended the idea to allow for multiple maps to be constructed: e.g., maps of foregrounds in addition to the CMB. This basic technology has been incorporated into the most recent results from Planck~\cite{planck18_ClLikelihood}. Groups are now applying the technology to galaxy surveys~\cite{Tsaprazi:2021mft,loureiroAlmanacfieldlevelWeaklensing2022,andrews_Bayesian3DFieldLevelPNG2022,KappaMassMapping_Fiedorowicz2022,MatrixfreeBayesian_Jasche2015,Bayesiandeepgalaxysurveys_Ramanah2019}.

One way to understand the advantage of the field-level approach is to return to the DES+SPT analysis: first CMB lensing maps were made using the traditional quadratic estimator~\cite{Hu:2001tn} and then they were used to construct two-point functions. However, the data in DES itself could in principle help improve the fidelity of the CMB lensing maps: after all, the deflection of the CMB photons is due (at least in part) to the very structure that DES measures. Combining this information would clearly create a better CMB lensing map. Using that improved map with DES maps though would be a form of double counting, so it makes sense to do everything at once: create all the maps and estimate all the power spectrum simultaneously. In the particular example of CMB lensing, the problem is not trivial but \citet{milleaBayesianDelensingOfCMB2019,MilleaBayesianDelensingDelight2020,milleaOptimalCMBLensingFieldlevel2021} have made significant progress simultaneously measuring the lensing field, the primordial CMB, and several parameters that determine the relevant power spectra.

Here we use simulated data sets on large scales to (i) develop the machinery that can handle real data; (ii) explore some of the basics of field-level analyses; and (iii) provide an example of how the field-level analyses can be used to stress-test $\Lambda$CDM. Our example is related to the work in \citet{eriksen_BayesianCompSepCl2008}, except that we attempt to separate the late-time Integrated Sachs-Wolfe (ISW) signal from the primordial CMB anisotropies. \citet{hangGalaxyClusteringDESI2020} constrained the ISW and lensing amplitudes using the two-point correlation between the DESI Legacy Survey and Planck temperature and lensing maps, where as we are interested in the full posterior distribution of both the parameters, two-point functions, and the maps. 

We begin in \S II by explaining some of the details; then in \S III, we analyze simulated CMB data assuming that it consists only of noise and CMB anisotropies. We recover some of the known problems of the maximum posterior solution (the Wiener filter) and show that these can be mitigated by instead using samples of the full posterior. Then, in \S IV, we introduce the ISW component and try to separate that from the primordial anisotropies. The degeneracies make this problematic at the map level, but the sampler produces an unbiased power spectrum for each. This is crucial, as the cosmological parameters themselves are embedded in the spectrum so if the spectrum is unbiased, then the parameters will be as well. Specifically, we introduce two free amplitudes of each component that multiply the fiducial spectra and show that the field-level analysis that simultaneously solves for the map values and the parameters produces unbiased estimates of the parameters.

The ensuing constraints on the amplitude of the ISW spectrum are not very restrictive, so in \S V, we explore the possibility of adding in other tracers, the galaxy density in several tomographic bins. This adds to the number of free parameters in the field-level analysis but we show that it produces a higher fidelity ISW map and a fairly tight constraint on the amplitude of the ISW spectrum. This leads to the prospect of stress testing $\Lambda$CDM by introducing amplitudes in front of all spectra (CMB lensing; galaxy density; cosmic shear) in addition to the standard cosmological parameters: a measurement in which any one of these amplitudes is determined to deviate from unity will disprove $\Lambda$CDM by demonstrating that structure does not grow in time as predicted by the model. 

In short, our goal in this paper is to explain (to some, much of this will not be new) what to expect when carrying out a field-level analysis; demonstrate how well it does on simulated data with increasing numbers of components and probes, and point the way to a simple but powerful way to stress test the fiducial cosmological model. We share our conclusions and thoughts about the next steps in \S VI.

\section{Theoretical framework}
\subsection{Field-level multiprobe analysis}
\label{sec:Framework}
\begin{figure}[!htbp]
\centering
\includegraphics[width=1\hsize]{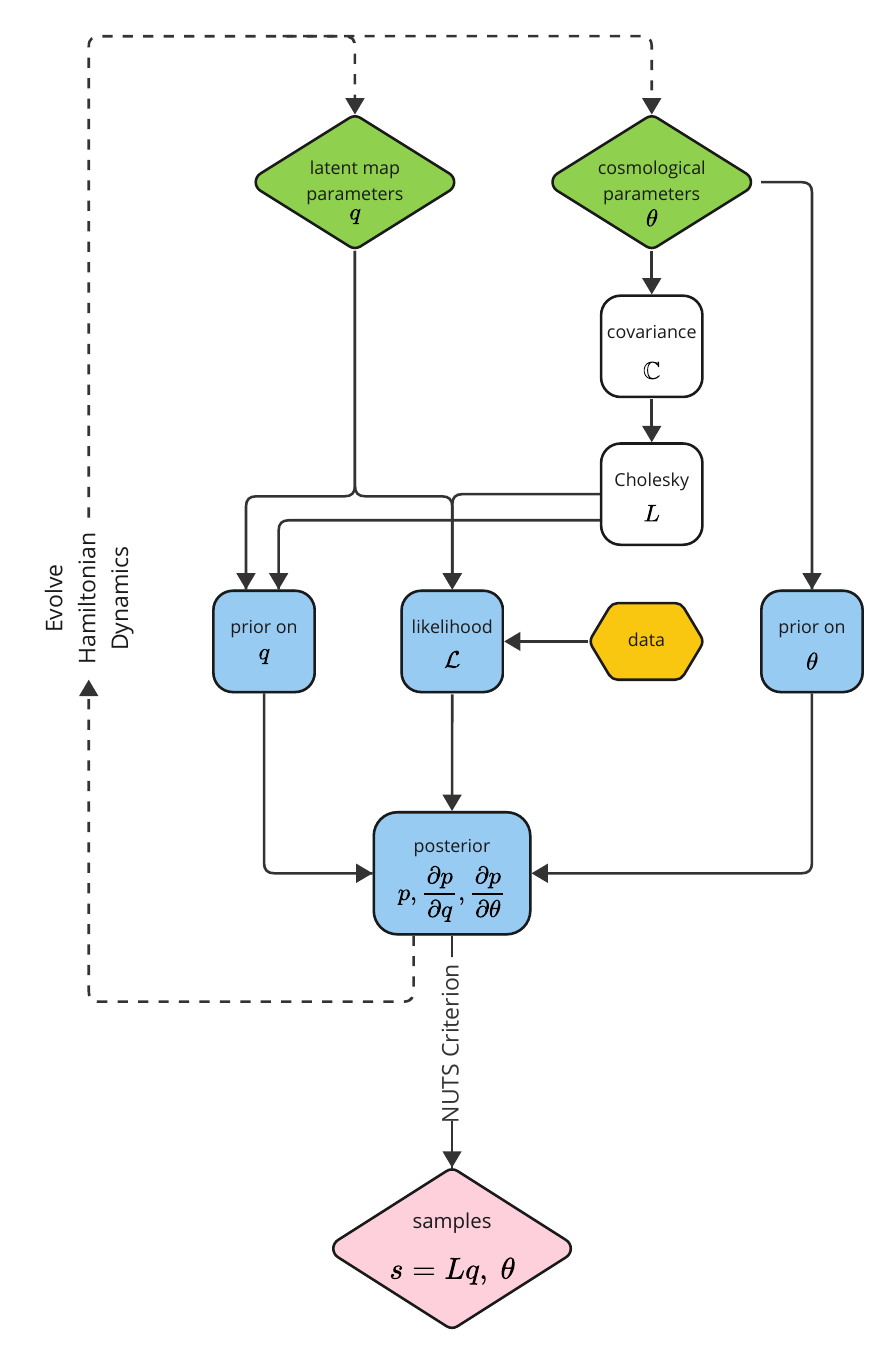} 
\caption{The flow chart of a general field-level multiprobe analysis that accumulates samples of both the cosmological parameters and the tracer maps (values of each signal in each pixel) as discussed in \S \ref{sec:Framework} and \S \ref{sec:Methodology}. We start from the prior distribution of the cosmological and latent map parameters. We then use the realized cosmological parameters to construct the covariance of the tracers, which in turn transforms the latent map parameters into physical tracer maps. The covariance, tracer maps, and observed data are then combined into the likelihood and -- after multiplying by the priors -- the posterior functions. If the above calculations are all programmatically differentiable, we can calculate the derivatives of the posterior function easily, and use HMC No-U-Turn Sampler (HMC-NUTS) to efficiently sample from the very high dimensional posterior space. In this diagram, diamonds denote sampled parameters, squares denote model-relevant functions, and the hexagon is the (fixed) observed data vector. The pink diamond represent samples that represents the posterior space.
}
\label{fig:flow_chart}
\end{figure}

The general problem of a field-level multiprobe inference is summarized in Fig.~\ref{fig:flow_chart}. The data is an observation, or a set of observations, on the sky. For concreteness, we will focus on the synergy between CMB experiments and photometric galaxy surveys, but the argument generalizes to any combinations of probes. 

The data is assumed to consist of a set of signals and noise:
\begin{equation}
\data = \sum_\alpha \signal^\alpha + \noise.
\end{equation}
Our model assumes that the signals $\signal^\alpha$ in the data are drawn from a Gaussian distribution with mean zero and covariance matrix, $\C^{\alpha\beta}(\theta)$, where $\theta$ represents cosmological and nuisance parameters. The noise is also drawn from a Gaussian distribution with mean zero and known covariance matrix $\C^n$.

The likelihood for obtaining the data given the cosmological parameters and the signals is
\begin{equation}
-2\ln\like = \left[\data-\sum_\alpha\signal^\alpha\right]\, [\C^n]^{-1} \,
\left[\data-\sum_\beta \signal^\beta\right] + \ldots
\end{equation}
where the additional terms are irrelevant, and the products on the right involve all pixels. That is, in the case of a single survey with $\npix$ pixels, $\data$ is a set of the values in all the pixels, and $\C^n$ is a $\npix\times\npix$ matrix. If only one signal contributes, then $\signal$ also has $\npix$ values; if more signals are assumed, then the total number of parameters in all the $\signal^\alpha$ will be $\npix\times N_{\rm signal}$. When data from multiple surveys are used, $\data$ will be a concatenated version of all the individual data sets and different signals can contribute to different data sets.

Using the Bayes theorem, we can invoke the prior on all the signals and the parameters. Since we are confining our analysis to large scales throughout, the prior on all signals is Gaussian, and the posterior is

\begin{align}
    \nonumber
    -2 \ln p = &-2 \ln\like + \sum_{\alpha\beta} \signal^\alpha \left(\C^{-1}(\theta)\right)^{\alpha\beta} \signal^\beta \\
    \label{eqn:post}
    &+ \ln \det{\C(\theta)} - 2\ \ln \mathrm{prior(\theta)}
\end{align}
where irrelevant terms have been dropped. The parameters in this posterior are $\theta$ (which determines $\C$) and the map(s) $\signal^\alpha$. For example, in the case of a single survey, if there is one signal contributing and there are 5 cosmological parameters, then the number of parameters we use to fit the $\npix$ data points is $\npix+5$. There are often cases where there are two or more signals contributing. For example, below we model the CMB as consisting of the signal from the last scattering surface plus the contribution from the late-time ISW effect. In that case, there will be $2\npix+5$ free parameters.

As described in \S II.C, we will draw samples from this posterior. The accumulated samples of both the maps and the cosmological and nuisance parameters are fully consistent in the Bayesian sense. More precisely, the distribution of the values of map pixels $\signal^\alpha$ will provide a set of posterior samples of the signals, and the distribution of the parameters will constrain the relevant models of interest. These distributions will be consistent with one another, so that for example in a sample with a large $\C^{\alpha\alpha}$, the signal $\signal^\alpha$ everywhere is likely to have a larger dispersion.

\subsection{Pixels}

Above we glossed over the details of the map. Here, we review the basics of pixels in terms of the coefficients of spherical harmonics and explain why we choose to work with this basis.

Consider a map on the curved sky $\signal(\bfn)$, where $\bfn$ is a 3-dimensional unit vector. Analogous to Fourier transformations in  Euclidean spaces, we can study this field in the frequency (or harmonics) space via forward and inverse spherical harmonics transform (SHT), 
\begin{align}
    \signal(\bfn) &= \sum_{lm}\signal\lm Y\lm(\bfn) \\
    \signal\lm &= \int \frac{d\Omega}{4\pi} \signal(\bfn) Y^*\lm(\bfn)
\end{align}
where $Y\lm(\bfn)$'s are the set of orthonormal spherical harmonics. We adopt the \healpix pixelization strategy (where the angular resolution is specified by a single parameter \nside), and use the discretized SHT as implemented by the \texttt{healpy} library \cite{gorskiHEALPix2005,zoncaHealpy2019}. As usual in cosmological analyses, we drop the monopole and dipole modes ($l=0,1$).

In general, if the field $\signal$ is statistically homogeneous and isotropic, it is more advantageous to study $\signal$'s correlation structure in harmonic space. In real space, the correlation function between two line-of-sight directions is given by 
\begin{equation}
    w_{\bfn,\bfn'} = w(|\bfn-\bfn'|) = \langle \signal(\bfn)\signal(\bfn') \rangle
\end{equation}
where we see that the correlation function has dense off-diagonals. For a discretized map with \nside\  resolution, the size of $w$ scales as $\nside^{4}$, which quickly becomes impossible to handle (for example, an $\nside=256$ map has an angular resolution of $27''$ and $8 \times 10^5$ pixels; the full pixel-pixel covariance matrix totals $5$ terabytes). 

However, $\signal$'s power spectrum ($\signal$'s correlation function in harmonic space), $\covs$,  defined by
\begin{equation}
    \langle \signal\lm^\alpha \signal_{l'm'}^{*,\beta} \rangle = 
    \delta_{ll'} \delta_{mm'} \covs_l^{\alpha\beta}
\end{equation}
is diagonal in this basis and depends only on the multiple moment $l$ and the different sets of signals assumed. Therefore, the amount of memory needed to manipulate $\covs$ is linear in \nside. One important caveat to this simplicity is that the field must be homogeneous and isotropic, and these assumptions fail in the presence of instrumental noise patterns, partial sky coverage, and masking. 

\subsection{Methodology}
\label{sec:Methodology}
Here we present the details of our implementation of Fig.~\ref{fig:flow_chart}.
The fundamental idea behind all MC sampling techniques is: start from the current sample; find the next point in the parameter space and generate a probabilistic proposal to make it a sample (both operations may involve repeated evaluations of the posterior density). How to find the next point and what proposal to make are algorithm-specific, however, they in general satisfy the principle of detailed balance such that, in the limit of large sample size, the samples approximate the posterior distribution. 

The efficiency of MC sampling rests on the suitability of the MC algorithm for the specific inference context and the effective computation of the posterior distribution. 

For the first point, since we are inferring both the map pixels and the cosmological parameters, the dimensionality of the posterior space will be quite large. For example, for our final analysis in \S V which includes 8 tracer maps at $\nside=32$, the total dimensionality of the posterior space is 73704. This is too large for traditional Monte Carlo techniques (such as Metropolis–Hastings) to operate efficiently. Intuitively, this is because as dimensionality increases, the ratio between the neighboring volume pointing towards and away from a particular point in the parameter space (e.g., the mode of the distribution) decays exponentially. Thus, the Random Walk Metropolis algorithm becomes overwhelmingly likely to propose samples outside the typical set, where the target density and hence the acceptance probability vanishes \cite{betancourtHMC}. 

Hamiltonian Monte Carlo (HMC) solves this efficiency problem in high dimensional spaces \cite{nealHMC2011,betancourtHMC,HoffmanNUTS_2011arXiv1111.4246H}. In the HMC framework, we augment the parameter space with a conjugate momentum space and use the gradient of the log posterior surface to guide us to sample only near the bulk of the probabilistic mass \cite{betancourtHMC,nealHMC2011}. In order to avoid traditional HMC's sensitivity to hyper-parameters such as the integration steps, we further employ the No-U-Turn Sampler (NUTS) variation of the HMC, first proposed by \citet{HoffmanNUTS_2011arXiv1111.4246H}. 

This leads us to the second point on computational efficiency. HMC samplers require repeated evaluations of the posterior function and its gradient. Since we want to develop a multiprobe field-level framework that easily extends to different observables and cosmological models, we do not want to hard-code the derivatives in advance. Instead, we choose to make the framework pragmatically differentiable through the JAX auto-differentiation library in python, which interfaces smoothly with the \texttt{numpyro} implementation of the NUTS \cite{binghamPyro2018,phannumpyro2019}.

Turning to the specific problem of the posterior function computation. We start at the top of Fig.~\ref{fig:flow_chart} and break this calculation into several parts:

\begin{itemize}
\item We start with a proposal for the cosmological parameters $\theta$ and the latent map parameters. The latent map parameters are a set of uncorrelated standard Gaussian variables $q$ which we will later transform into the signal maps. 
\item Calculate $\C$ given the cosmological parameters
\item Transform the latent map parameters using the Cholesky decomposition ($\C=LL^t$) of the covariance matrix:
$\signal=L r$. (The prior of the maps becomes $s^2/(2\C) = q^2/(2\mathbb{I})$).
\item Combine the maps $s$ with the data to calculate the likelihood (forward modeling) 
\item Use the likelihood and the prior to calculate the posterior and its derivative with respect to the parameters
\item If the NUT criterion is satisfied, accept this as a valid sample
\item Use the leapfrog method to generate another sample
\end{itemize}
In practice, through the JAX auto-differentiation library, this framework provides information on both the posterior and its gradient. 

The treatment of the covariance function deserves more discussion. In this paper, we will keep the shape of the spectra fixed and allow for free amplitudes $(A^\alpha)^2$. We assume the amplitudes have fiducial values equal to one and have a uniform prior distribution. We then construct the full covariance matrix $\covs^{\alpha\beta}$, which consists of the auto- and cross-spectra of each signal. 
In principle, since $\C$ encodes the covariance between all the pixels for all the tracer maps, its dimensionality is very high. For a single \healpix map at the resolution of $\nside$ (or a limiting resolution of $\lmax = 3\nside-1$), there are 
\begin{equation}
    \sum_{l=2}^{\lmax} (2l+1) = \lmax^2 + 2 \lmax -3
\end{equation}
degrees of freedom ignoring monopole and dipole modes. Again, for our final analysis in \S V, which includes 8 tracers, the size of $\C$ is on the order of $8^2 \times \lmax^4$. The efficient computation of this covariance matrix is one of the limiting factors in the feasibility of field-level analysis. However, in the limit of full sky and when all the fields are homogeneous and isotropic, the sub-block of $\C$ for each tracer is diagonal in the $a\lm$ basis. Thus, we can bring $\C$ into block diagonal forms, with $\lmax - 2$ unique $\C_l$ sub-blocks on the diagonal. Each $\C_l$ sub-block has size $8 \times 8$, describing the correlation between the 8 tracers at mode $l$. Looking ahead, we will be considering the primordial CMB signal (modulated by $\AP$); the late-time ISW effect (modulated by $\AI$); and the galaxy density in 6 tomographic bins (modulated by $b_i$'s). The first of these is uncorrelated with the rest, so the ensuing $8\times8$ sub-block matrix will be 
\begin{align}
    &\C_l(b_1,...,b_6,\AI,\AP) = \notag  \\
    &\begin{pmatrix}
    b_1^2\C^{\G_1,\G_1}_l    & b_1 b_2 \C^{\G_1,\G_2}_l & ... & b_1 \AI \C^{\G_1,\I}_l & 0\\
    b_1 b_2 \C^{\G_2,\G_1}_l & b_2^2 \C^{\G_2,\G_2}_l   & ... & b_2 \AI \C^{\G_2,\I}_l& 0\\
    ... & ... & ... & ... \\
    b_1 b_6 \C^{\G_6,\G_1}_l & b_2 b_6 \C^{\G_6,\G_2}_l & ... & b_6 \AI \C^{\G_6,\I}_l& 0\\
    b_1 \AI \C^{\I,\G_1}_l  & b_6 \AI \C^{\I,\G_2}_l  & ... & (\AI)^2 \C^{\I,\I}_l& 0\\
    0&0&\ldots & 0&  (\AP)^2 \C^{\P,\P}_l\\
    \end{pmatrix}\eql{offdiag}
\end{align}

With this computationally efficient representation of the covariance matrix, we can transform the latent map variables into the proper tracer maps through either the sub-blocks' Cholesky representations or their eigen-decomposition. We experimented with both, and found the former to be an order of magnitude faster (see also \citet{loureiroAlmanacfieldlevelWeaklensing2022}). In summary, the algorithm is very fast. For the largest model we considered in \S \ref{sec:CMBISWGal}, the analysis was done on an Apple M1 chip running overnight. 

\section{Reconstruction of the primordial CMB map}
\label{section:CMB}

We start with the simplest possible example, a single simulated CMB all-sky map. Although simple, this model demonstrates the key behaviors of two ways of using the posterior: identifying the free parameters by finding the point at which the posterior is maximum (hereafter, maximum a posteriori or MAP) and generating samples of the posterior (hereafter sampling). We compare the potential biases of both methods and discuss the implication for the field, two-point, and cosmological parameter constraints. In some ways, the idea of asking whether an estimator is biased is introducing frequentist ideas into a Bayesian discussion. Nonetheless, we think that understanding these biases is an important step towards the ultimate goal of extracting the correct cosmological parameters from the data. The intuition we find here will serve us well in the subsequent more complex cases.

The primordial CMB temperature fluctuations $\sP$ originate from the time of recombination ($z_*\approx1100$), when the photons decoupled from the photon-electron-proton fluid as the universe cooled below a few percent of the ionization energy of hydrogen. We assume (and all current data is consistent with this assumption, with the tightest constraints coming from the ~\citet{https://doi.org/10.48550/arxiv.1905.05697})  that the resulting temperature variation is a homogeneous and isotropic random Gaussian field which is fully characterized by the power spectrum $\CP_l$. Throughout this paper, we assume a fiducial cosmology of $H_0 = 100h = 67.5\text{km/Mpc/sec}, \quad \Omega_b h^2 = 0.0219, \quad \Omega_c h^2 = 0.1139, \quad A_s = 2 \times 10^{-9}$ and $n_s = 0.965$. 

\subsection{Biases of the optimal estimator}
\subsubsection{Fixed cosmological parameters}
The observed temperature data $\dT(\bfn)$ is the superposition of the primordial field $\sP(\bfn)$ and noise $\nT(\bfn)$. 
\begin{equation}
    \label{eqn:CMBmodel_model}
    \dT(\bfn) = \sP(\bfn)+\nT(\bfn)
\end{equation} We call this the CMB model. We ask: Given $\dT$ and perfect knowledge of $\sP$'s and $\nT$'s theoretical power spectra ($\CP$ and $\CNT$ respectively), how well can we reconstruct the primordial field? Additionally, how accurate is the power spectrum of the reconstructed field?

In the Bayesian framework, the posterior probability in Eq.~\ref{eqn:post} reduces to
\begin{equation}
    \label{eqn:CMBmodel_post}
    -2 \ln p(\sP|\dT) \propto \sum\lm \left( \frac{|\dT\lm-\sP\lm|^2}{\CNT_l} + \frac{|\sP\lm|^2}{\CP_l} \right)
\end{equation}
where we drop the determinant terms since the CMB and noise spectra are assumed known and fixed. The MAP solution for $\sP$ is then given by the Wiener filter 
\begin{equation}
    \label{eqn:MAP_s_noamp}
    \sPMAP\lm = \frac{\CP_l}{\CP_l+\CNT_l}\dT\lm.
\end{equation}
The mean power spectrum of the MAP estimator
\begin{equation}
    \label{eqn:MAP_C_noamp}
    \langle\CPMAP_l \rangle\equiv \langle \frac{1}{2l+1}\, \sum_m \vert \sPMAP\lm \vert^2 \rangle= \frac{\CP_l}{\CP_l+\CNT_l}\CP_l
\end{equation}
is known to be biased~\cite{1992ApJ...398..169R,kostic2022consistency}, an effect more prominent in the low signal-to-noise ratio (SNR) regime. 

\begin{figure}[!htbp]
\centering
\includegraphics[width=1\hsize]{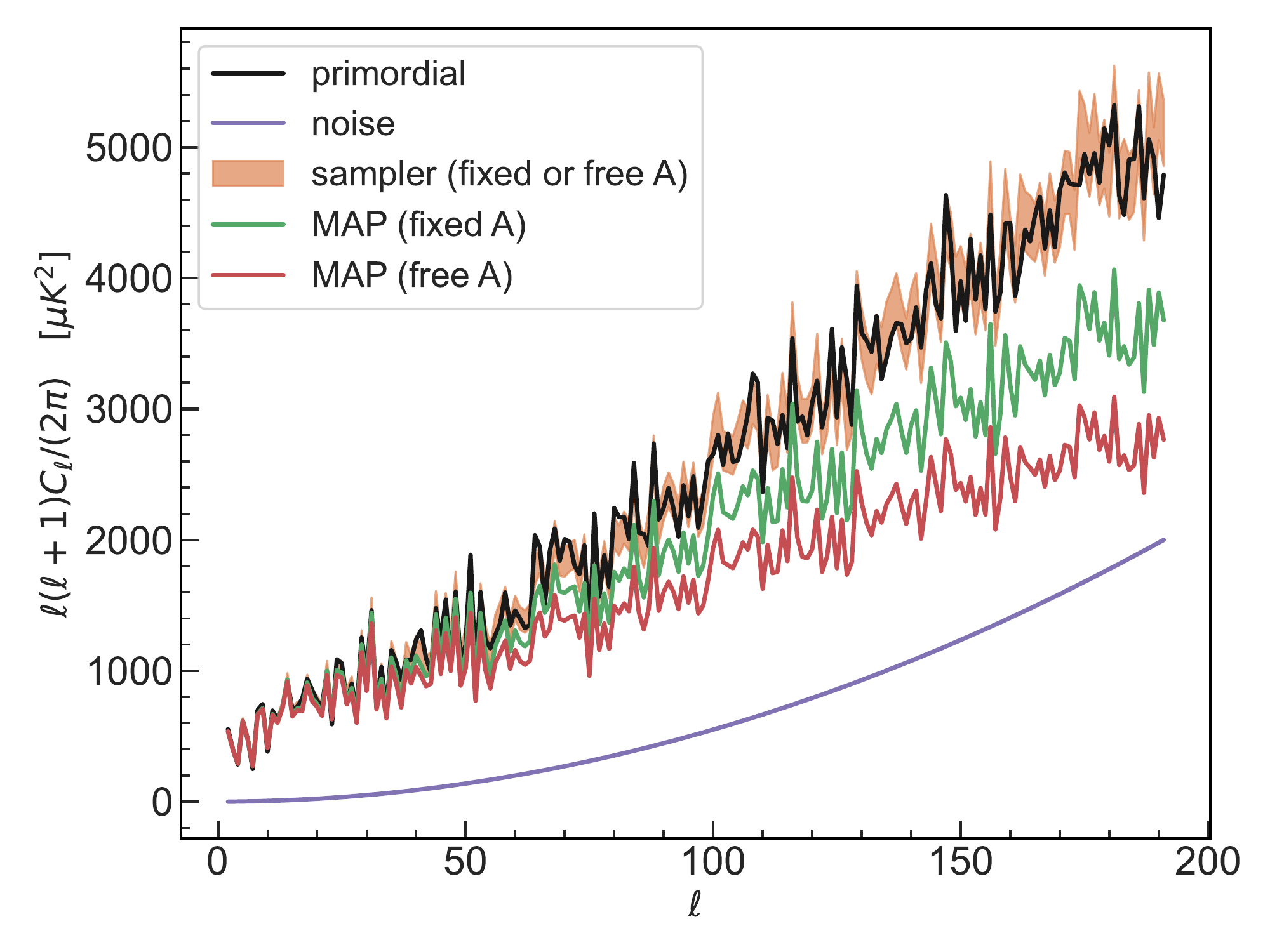} 
\caption{A numerical experiment of CMB reconstruction that demonstrates the difference between the MAP estimators and the sampling-based estimators. We generated the truth map from a fiducial $\CP$ and added isotropic noise. The truth and noise spectra are shown in black and purple. We test two reconstruction models, the first has a fixed $\CP$ (Eq.~\ref{eqn:CMBmodel_post}), and the second has a free amplitude $A^2$ modulating the fiducial $\CP$. Their MAP solutions are given by Eqs.~\ref{eqn:MAP_C_noamp} and \ref{eqn:MAP_C_amp}, and are shown in green and red respectively. We confirm the analytical solutions match the results from direct numerical optimization. Both two-point and MAP estimators are biased lower than truth and the free amplitude model has a greater bias. We then sampled $\sP$ directly from the posterior space. The spectra of the mean maps for the fixed and free amplitude cases are biased (both overlap the green curve). However, in both cases, the distributions (the orange shaded region represent $1\sigma$ credible interval) of the spectra scatter around the truth and are unbiased. }
\label{fig:CMB_summary}
\end{figure}

To implement, we simulate $\sP$ from the fiducial cosmology power spectrum on a \healpix grid of $\nside=64$ and then inject isotropic white noise $\nT$ with a relatively high variance of $\Var(\nT) = 1000 \mu K^2$. Even though this exceeds noise in Planck by several orders of magnitude, we use this value to demonstrate the difficulties of extracting the signal in the presence of appreciable noise. The power spectra of the truth map (black) and the recovered MAP map (green) are shown in Fig.~\ref{fig:CMB_summary}. As the amplitude of the noise spectrum (purple) rises on small scales, the MAP spectrum is increasingly suppressed. On the field level, this means that $\sPMAP\lm$ is damped for over and underdensities on scales that have small SNR. The estimator does not have an additive bias but does have a multiplicative bias, i.e., $\langle \sPMAP\lm/\sP\lm \rangle = \CP_l/(\CP_l+\CNT_l)$ for both the real and the imaginary components. This is shown in the top panel of Fig.~\ref{fig:CMB_scatter}. 

\subsubsection{Varying cosmological parameters}
In real cosmological analyses, we are also interested in cosmological parameters (such as the primordial amplitude, spectral index, etc.) that modify the shape and amplitude of the power spectrum. We want to know how the field, two-point, and parameter MAP estimators behave when the spectrum is allowed to change.

For example, consider modulating the fiducial power spectrum $\CP$ with a scale-invariant amplitude $A^2$, where $A$ has a flat prior. The new MAP solutions are given by
\begin{align}
\label{eqn:MAP_s_amp}
\sPMAP\lm &= \frac{(\AMAP)^2 \CP_l}{(\AMAP)^2  \CP_l+\CNT_l}\dT\lm \\
\label{eqn:MAP_C_amp}
\CPMAP_l &= \frac{(\AMAP)^2\CP_l}{(\AMAP)^2\CP_l+\CNT_l}(\AMAP)^2\CP_l
\end{align}
and $\AMAP$ satisfies
\begin{equation}
    \sum_{l} (2 l + 1) \left\{ 1- \frac{(\AMAP)^2 \mathbb{D}_l \CP_l}{((\AMAP)^2 \CP_l + \CNT_l)^2}\right\} = 0
\end{equation}
where $\mathbb{D}_l$ is the data power spectrum. 

\begin{figure}[!htbp]
\centering
\includegraphics[width=1\hsize]{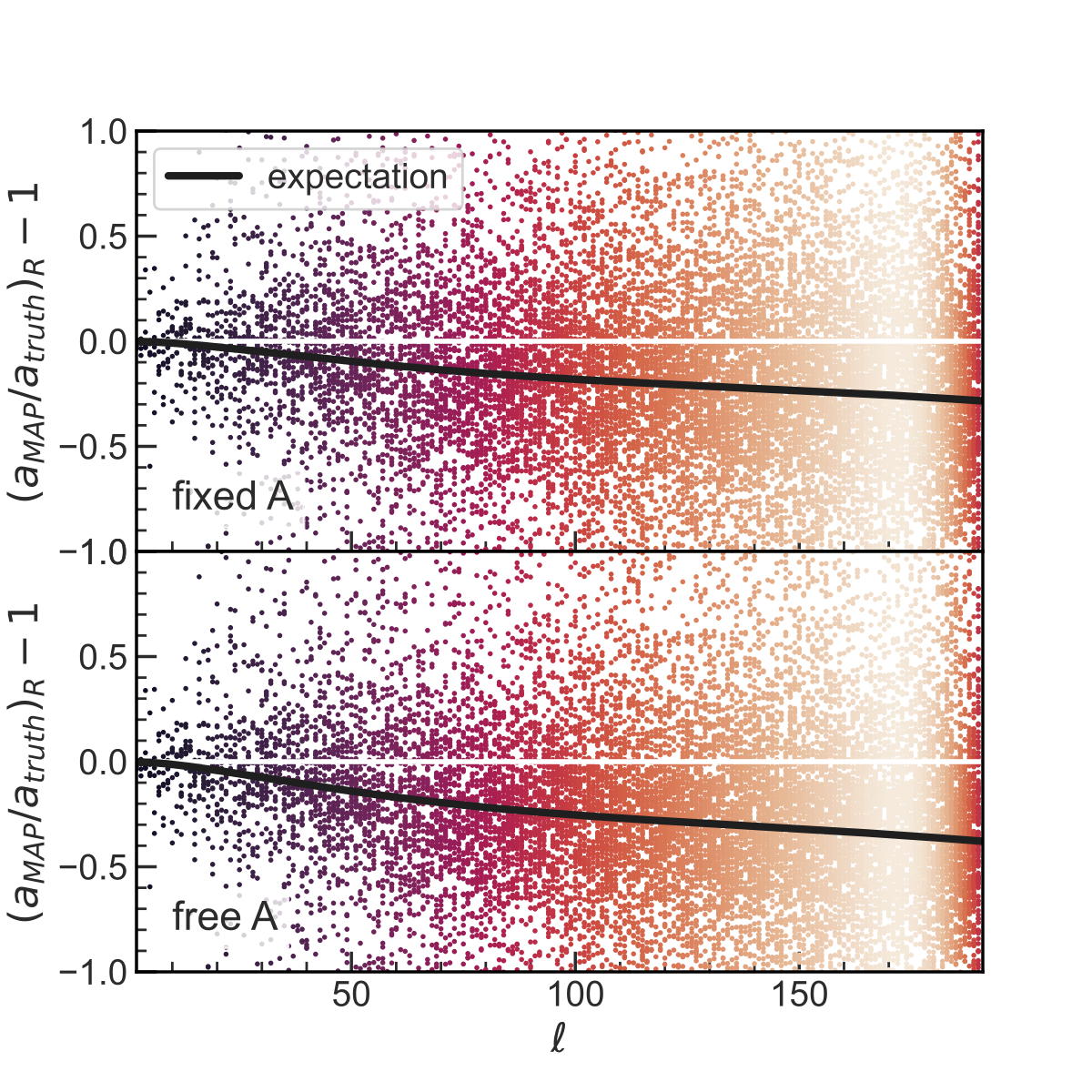} 
\caption{The scatter plot of the ratio between the real components of the MAP $a\lm$ and that of the truth $a\lm$ as a function of $l$. The cases for fixed and varying amplitude parameters are shown in the upper and lower panels respectively. In each case, the mean of the ratios (black curve) matches the Wiener filter expectations (with the MAP amplitude in the case of varying amplitude). The color map represents point cloud density.}
\label{fig:CMB_scatter}
\end{figure}

When noise is present, $\AMAP < 1$, in this case equal to $0.81$ (see Fig.~\rf{CMB_A}).  Thus, by comparing Eqs.~\ref{eqn:MAP_C_noamp} and \ref{eqn:MAP_C_amp}, we see that the new MAP is biased even lower than the truth. We can apply this model to the same reconstruction experiment as before. The result for the new power spectrum estimator is shown in red in Fig.~\ref{fig:CMB_summary}. The multiplicative bias in the field-level estimator is shown in the bottom panel of Fig.~\ref{fig:CMB_scatter}.

For power spectra with complicated parameter dependence, we often lack analytical optimal solutions. However, qualitatively speaking, if an increase in the parameter increases the amplitude of the spectrum as in this case, then the parameter will be underestimated by optimal inference, and vice versa. 

\subsection{Sampling the CMB field}
\label{sec:CMB_SamplingtheCMBfield}
\subsubsection{Fixed cosmological parameters}
Now we seek an unbiased estimator for $\CP$ that also has a convenient notion of uncertainty. Let us again first fix $A = 1$ and draw a sample of maps $\{\sP_{lm,i}\}_{i=1,...,N}$ directly from the posterior distribution (Eq.~\ref{eqn:CMBmodel_post}). Using this set of maps, we can construct an associated set of power-spectra samples
\begin{equation}
    \{\CP_{l,i}\} = \{ \sum_m \frac{1}{2l+1} |\sP_{lm,i}|^2 \}
\end{equation}
Let us call the ensemble average of $\{\sP_{lm,i}\}$ and $\{\CP_{l,i}\}$ as $\mean{\sP}$ and $\mean{\CP}$, respectively. It is crucially important that the power spectrum of $\mean{\sP}$ is \emph{different} from $\mean{\CP}$. 

We claim, in the limit of sufficient sample size $N$,
\begin{enumerate}
    \item $\mean{\sP}$ and its spectrum are exactly the field and two-point MAP estimators (Eqs.\ref{eqn:MAP_s_noamp} and \ref{eqn:MAP_C_noamp}), and they have the Wiener filter multiplicative bias. 
    \item On the two-point level, the samples $\{\CP_{l,i}\}$ give a proper Bayesian credible interval centered around the truth. 
    \item Further, $\mean{\CP}$ (and more generally, the mean of any $n$-point power spectrum samples) is an \emph{unbiased} estimator in the frequentist sense (when we have multiple data realizations).
\end{enumerate}
We prove these claims in Appendix \ref{appendix:proof}. However, intuitively, how can the power spectrum of the mean map be biased while the sampled power spectrum be unbiased? One way to understand this is to think of each sampled map as $\sP = \sPMAP + s'$. When we compute the power spectrum $\langle |\sP|^2 \rangle$, the $\langle |s'|^2 \rangle$ term exactly compensates for the deficiency of the MAP spectrum. Alternatively, the $\sP$'s are normally distributed, and for any Gaussian distribution the mean is equal to the maximum, so $\mean{\sP}$ is the MAP solution. However, the power spectrum is not normally distributed; its expected value is an unbiased estimator of truth, and not the biased MAP solution. 

We continue with the numerical experiment above. This time, we construct an HMC NUT sampler following the prescription of \S \ref{sec:Methodology}, using Eq.~\ref{eqn:CMBmodel_post} as our posterior distribution. After the chain equilibrates, we draw 3000 $\sP$'s from the posterior space. We confirm that the power spectrum of the mean map exactly follows the MAP solution for the case of fixed parameters (the green curve in Fig.~\ref{fig:CMB_summary}). We further show the distribution of the sampled power spectra $\{\CP_{l,i}\}$ in Fig.~\ref{fig:CMB_summary} in shaded orange. Indeed, the distribution of the spectra covers the truth power spectrum within uncertainty, and $\mean{\CP}$ is unbiased. 

We note that similar phenomena have been observed in previous studies. For example, in Fig.~8 of \cite{MilleaBayesianDelensingDelight2020}, the authors find that the distribution of the sampled CMB spectra scatter around truth while the spectrum of the mean map is biased lower at small scales.

The above observations have the following implications. One must debias the sampled maps before using them for cosmological analysis, similar to how we currently correct for MAP maps (e.g., with analytical or Monte Carlo-based corrections). However, if we are only performing analysis on the two-point level, the samples are \emph{unbiased} and their distribution constitutes a convenient measure of uncertainty. In short, by considering the sampled power spectra, we recapture an unbiased estimator of $\CP$.

\subsubsection{Varying cosmological parameters}
The exact two-point statistics recovery motivates us to ask whether the sampled cosmological parameters that modify the power spectrum are also unbiased. To answer this question, we use the HMC NUT sampler from the previous section with an additional $A^2$ modulating the fiducial spectrum. We assume $A$ has a flat and wide prior on $[0.2,10]$ and collect 3000 samples after appropriate burn-in. 

\begin{figure}[!htbp]
\centering
\includegraphics[width=1\hsize]{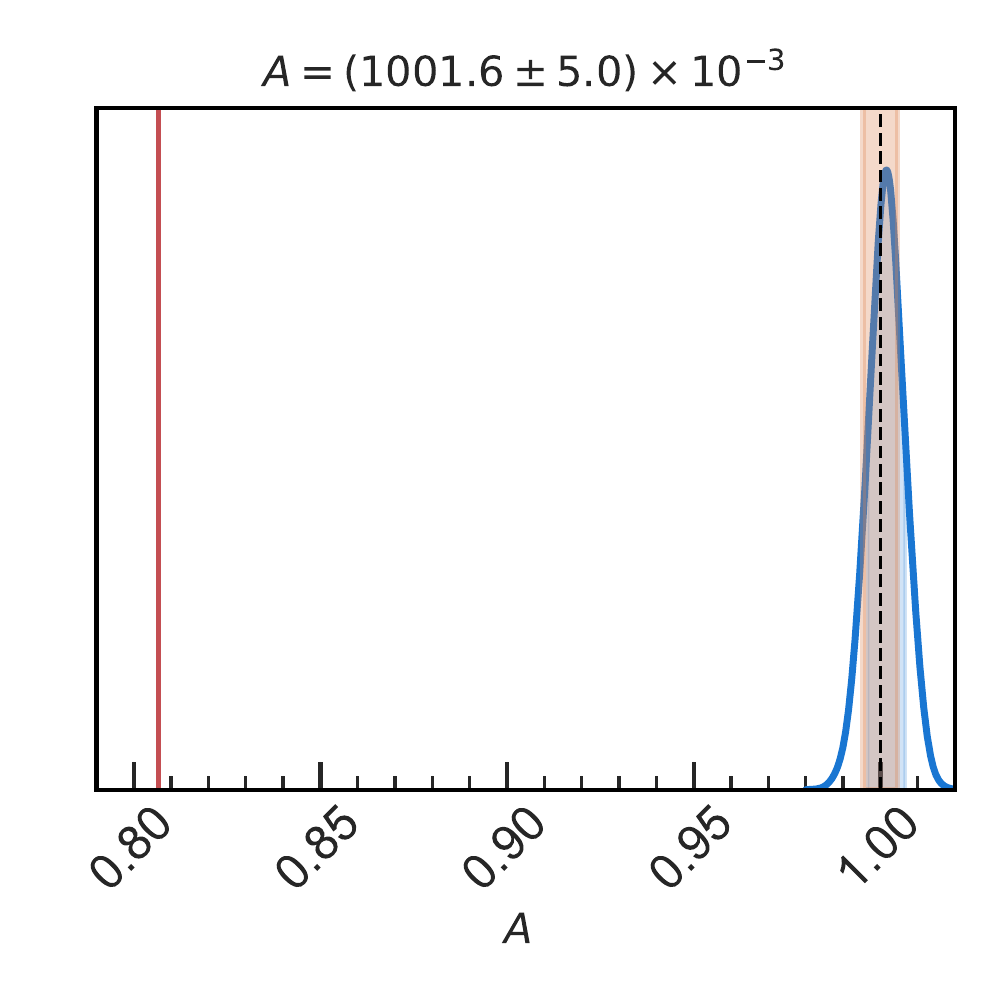} 
\caption{The posterior distribution of the amplitude $A$ (blue curve) compared to truth (black dashed line). The shaded orange area represents the Fisher forecast of the $1\sigma$ uncertainty centered on truth, and the shaded blue area represents the $1\sigma$ uncertainty reported by the sampler. The red line shows the MAP value for $A$, which is quite far from truth.}
\label{fig:CMB_A}
\end{figure}

We find that $A$ is unbiased with its marginal distribution shown in Fig.~\ref{fig:CMB_A}. We can also estimate the variance of its distribution, which is predicted by the inverse of the Fisher information 
\begin{equation}
\label{eqn:CMB_Fisher}
    \mathcal{F}^{-1} = \sum_l\frac{2(2l+1)}{1+(\CNT / \CP)^2},
\end{equation}
shown as shaded orange in the same figure. 

Since the expectation of $A$ is unbiased, it follows that the posterior samples $\{\CP_{l,i}\}$ again scatter around truth unbiased. In fact, their distribution overlaps that of the fixed amplitude model almost exactly as shown in Fig.~\ref{fig:CMB_summary} in shaded orange. Further, the power spectrum of the mean map is also not the MAP anymore; it is the MAP solution of the model with fixed cosmological parameter (Eq.~\ref{eqn:MAP_C_noamp}), as if $A$ is fixed to $1$. 

The important takeaway is the following. The MAP map and the mean sampled map are biased both on the field level and on the two-point level. However, the distribution of the sampled power spectra (and cosmological parameters that modulate them) is unbiased.

\section{Joint reconstruction of the primordial CMB and the ISW effect}
\label{sec:CMBISW}
Now, we expand on the CMB model and consider extracting the primordial and the ISW contributions from a single noisy temperature measurement. As we shall see, the MAP estimators for both signals are again biased on the field and the two-point level. The sampled fields have a multiplicative bias but their two-point statistics are unbiased. The new challenge in this case study is the field-level degeneracy between the primordial and the ISW maps, which motivates the multiprobe approach presented in the next section \cite{eriksen_BayesianCompSepCl2008}. We discuss the key properties of this degeneracy, which we expect to be quite general when one separates a low SNR map from a measurement based on a likelihood approach.

\subsection{The ISW effect}
\label{section:ISWtheory}
In the late universe, the primordial CMB fluctuations are modified by the ISW effect on very large scales~\cite{1985SvAL...11..271K,Crittenden_1996,
Fosalba:2003iy,Boughn:2003yz,scranton2003SDSS}. A photon is blue-shifted when descending a gravitational potential well and red-shifted when it escapes. When the universe began its accelerated expansion (in the dark energy-dominated era), the large-scale potential wells decayed. As a result, a photon will leave the decaying well (barrier) with more (less) energy than it enters. The observed ISW temperature modification is thus the integrated effect of the decaying potential well along the line of sight, and its 2-dimensional field is given by~\cite{Afshordi_2004,manzottiMappingISW2014}
\begin{equation}
    \label{eqn:2dISW1}
    A^{\mathcal{I}}(\bfn) = \int_0^\infty \frac{\partial \Phi(\bfx[\chi,\bfn],t(\chi))}{\partial t} \frac{2 e^{-\tau(\chi)}}{1+z(\chi)} d\chi
\end{equation} 
Here, $\tau(\chi)$ is the optical depth out to distance $\chi$ and $\Phi$ is the 3-dimensional gravitational potential field, which ultimately depends on the matter overdensities $\delta_m(\bfx[\chi,\bfn],t(\chi))$. By rewriting $\Phi$ in terms of $\delta_m$, and moving to the Fourier space, we can rewrite Eq.~\ref{eqn:2dISW1} as 
\begin{equation}
    \label{eqn:2dISW2}
    a\lm^{\mathcal{I}} = 4\pi i^l \int \frac{d^3k}{(2\pi)^3} I^{\mathcal{I}}_l(k) Y^*\lm(\bfk) \delta_{m}(\bfk,t_0)
\end{equation}
where 
\begin{equation}
    I^{\mathcal{I}}_l(k) = \int d\chi D(\chi) W^\mathcal{I} (k,\chi)j_l(k\chi)
 \label{eqn:window}\end{equation}
and the window function is given by
\begin{equation}
    \label{eqn:ISWwindow}
    W^{\mathcal{I}}(k,\chi) = -\Theta(\chi_*-\chi) \frac{3\Omega_m H_0^2}{k^2} \frac{\partial \ln((1+z)D(z))}{\partial t}
\end{equation}
where $D(z)$ is the growth function normalized to unity at $z=0$ and we approximate $\tau$ as zero through the epoch that the ISW is generated.

The forms of Eqs.~\ref{eqn:2dISW2} - \ref{eqn:ISWwindow} are not peculiar to the ISW effect - by modifying the window function $W(\chi,k)$ appropriately, the 2-dimensional observable of most tracers can be computed as a line-of-sight integral of $\delta_m$. For general tracers, $A$ and $B$, of this form, the covariance is
\begin{equation}
    \label{eq:Cab}
    \C^{A,B}_l = \frac{2}{\pi}\int_0^\infty k^2 dk P(k) I^A_l(k) I^B_l(k).
\end{equation}
One way of quantifying the correlation between different probes is to compute the scale-dependent correlation coefficients, defined as
\begin{equation}
    \label{eqn:rhoab}
    \rho^{AB} = \frac{\C^{AB}}{\sqrt{\C^{AA} \C^{BB}}}.
\end{equation}

\begin{figure}[!htbp]
\centering
\includegraphics[width=1\hsize]{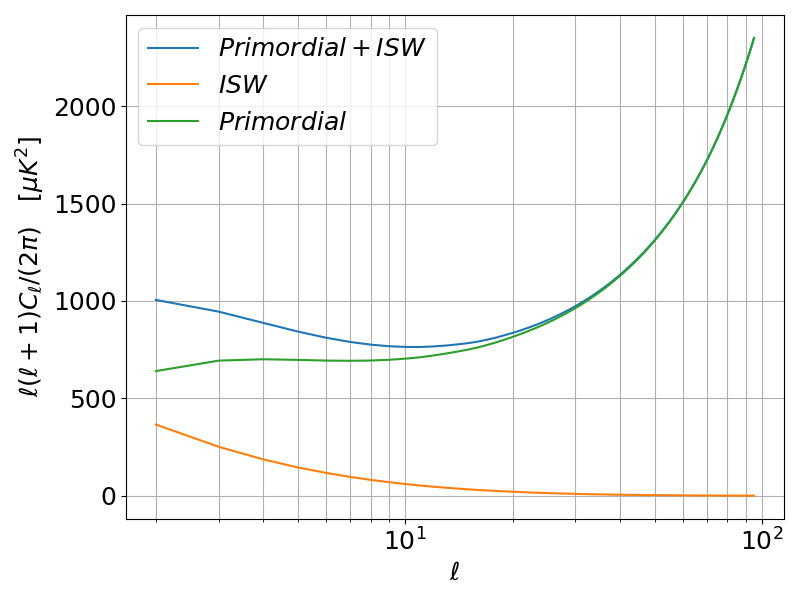} 
\caption{The power spectra of the primordial CMB and the ISW effect on large scales (low-$\ell$ modes). }
\label{fig:Cl}
\end{figure}

We again assume both fields are statistically homogeneous and isotropic, so the covariance is diagonal in the $a\lm$ basis. For example, the primordial CMB and the ISW effect are 
spatially independent ($\C^{\mathcal{I},\mathcal{P}} = 0$), and their autopower spectra are shown in Fig.~\ref{fig:Cl} ($\C^{\mathcal{I},\mathcal{P}} = 0$). The ISW signal is primarily confined to the very large scales ($l < 10$) that enter the horizon during the dark energy-dominated era. The ISW signal is also subdominant to the primordial signal on all scales, making it particularly challenging to reconstruct. 
We explore the case of nondiagonal covariance in \S \ref{sec:CMBISWGal} in the context of multiprobe joint reconstruction.

\subsection{Separating the primordial and ISW signals}
\label{sec:CMBISW_SeparatingthePIsignals}
The observed temperature data $\dT(\bfn)$ is the sum of the primordial field $\sP(\bfn)$, the ISW field $\sI(\bfn)$, and noise $\nT(\bfn)$, so the posterior distribution
\begin{align}
    &\ln p(\sP, \sI, A^P, A^I|\dT) \propto \notag \\
      &\sum\lm \half \left(\log {\AP}^2\CP_l + \log {\AI}^2\CI_l \right) \notag \\
      \label{eqn:CMBISWmodel_post}
    + &\sum\lm \left(\frac{|\dT\lm-\sP\lm-\sI\lm|^2}{2\CNT_l}
    + \frac{|\sP\lm|^2}{2{\AP}^2\CP_l}
    + \frac{|\sI\lm|^2}{2{\AI}^2\CI_l} \right)
\end{align}
We will refer to this model as the CMB-ISW model. 

One crucial difference between the CMB and the CMB-ISW model is that although both are constrained by the same amount of data $\dT$, the dimensionality of the latter's posterior space ($\sP$ and $\sI$) is (ignoring the cosmological parameters) twice than that of the former's (only $\sP$). In other words, if we have $N$ independent modes on the full sky, we are trying to constrain $2N$ parameters with $N$ data points in the CMB-ISW model. Hence, we expect significant degeneracy in the inferred $\sP$ and $\sI$ maps, and we want to explore how this affects the MAP and the sampling-based field-level reconstructions. We emphasize that this problem will be quite common in any field-level analysis where we wish to separate the different physical components of a single observed field. 

We will again tackle this problem in two ways, first by constructing the MAP estimators and then by sampling directly from the posterior distribution. 

\subsubsection{Fixed cosmological parameters}
Let us first fix $\AP = \AI = 1$ and seek the field-level MAP solutions. When we are trying to find the MAP of $\sP$, the effective noise is the sum of the instrumental noise and the ISW temperature fluctuation (and analogously for $\sI$). Thus, invoking the Wiener filter (Eq.~\ref{eqn:MAP_s_noamp}),
\begin{align}
    \label{eqn:MAP_s_noamp_CMBISW_P}
    \sPMAP\lm &= \frac{\CP_l}{\CP_l+\CNT+\CI}\dT\lm \\
    \label{eqn:MAP_s_noamp_CMBISW_I}
    \sIMAP\lm &= \frac{\CI_l}{\CI_l+\CNT+\CP}\dT\lm 
\end{align}
and similarly for their power spectra. 

\begin{figure*}[!htbp]
\centering
\includegraphics[width=0.7\hsize]{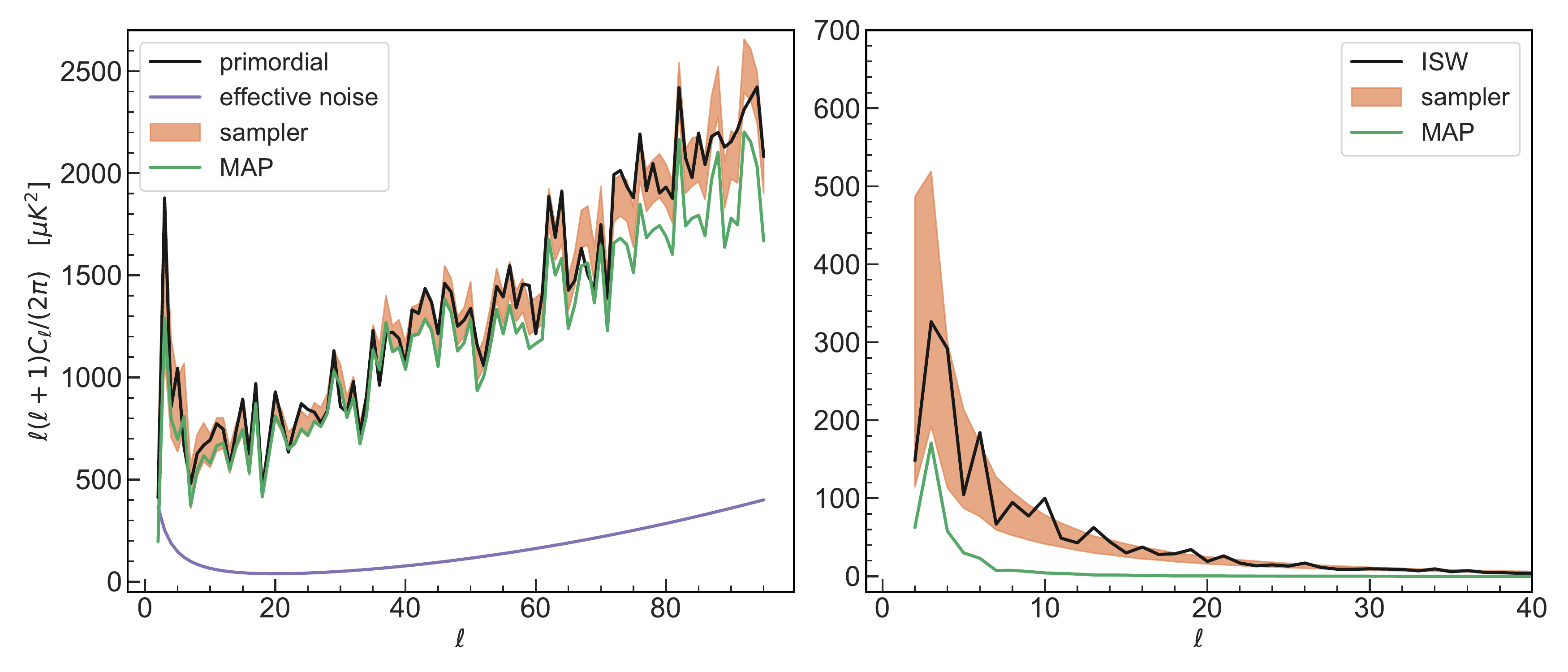} 
\includegraphics[width=0.7\hsize]{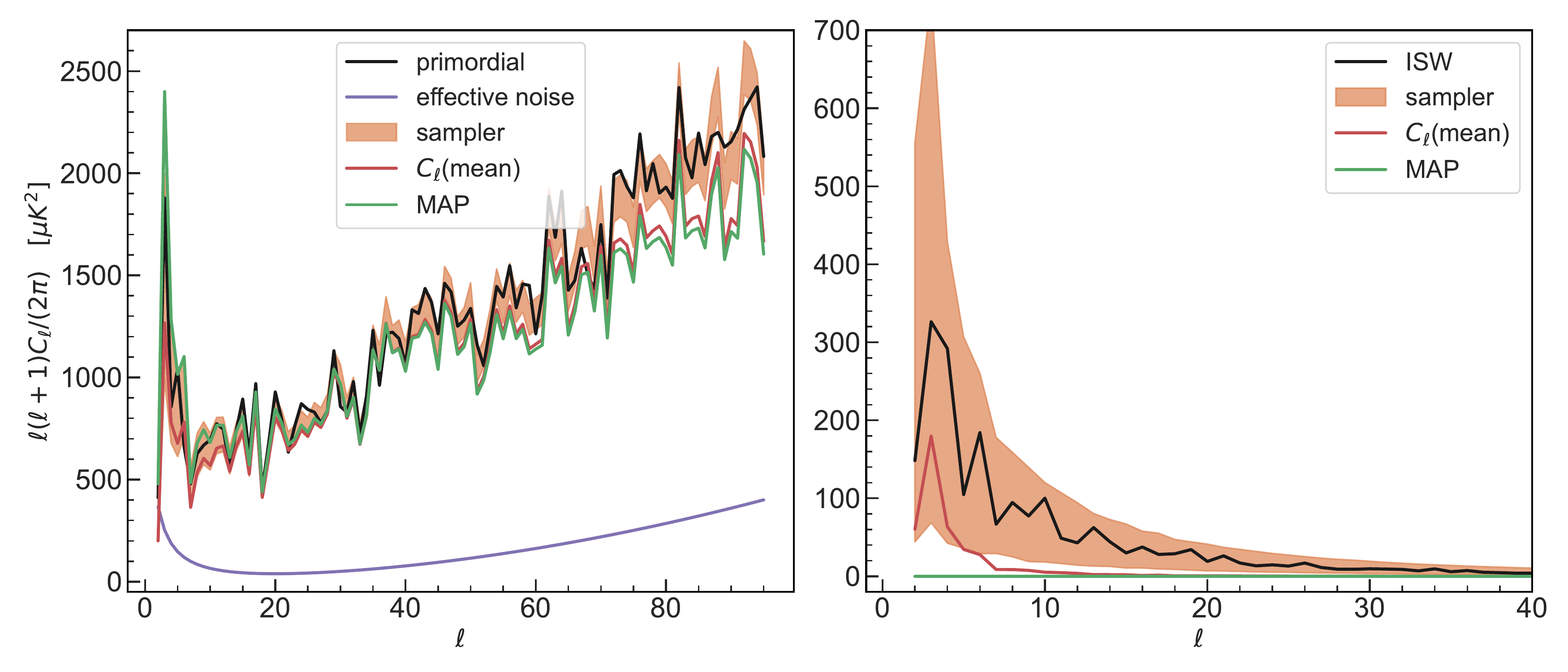} 
\caption{The recovered primordial ($\sP$) and ISW ($\sI$) spectra using the various models in \S \ref{sec:CMBISW_SeparatingthePIsignals}. The simulation is drawn from the fiducial cosmology and a noise variance of $200~\mu K^2$. The top and bottom panels show the case of fixed and free amplitude respectively, and the left and right panels show the primordial and the ISW results, respectively. For each case, the truth spectrum is shown in black, the MAP result is shown in green, and the $1\sigma$ credible interval of the sampled spectra is shown in shaded orange. The effective noise of the primordial maps is $\CI+\CNT$ shown in purple. The effective noise of the ISW map ($\CP+\CNT$) is above the ISW spectra in the right panels on all scales and hence not shown in the plot. Notice that in the case of free amplitude, the MAP solution is different from the $\C_l$ of the sampled maps (red).}
\label{fig:CMBISW_cl}
\end{figure*}

As in \S \ref{section:CMB}, we simulate this reconstruction method numerically by generating $\sP$ and $\sI$ with fiducial cosmology a \healpix grid of $\nside=32$ with $\Var(\nT) = 200~\mu K^2$. In the top panels of Fig.~\ref{fig:CMBISW_cl}, the truth power spectra are shown in black, the MAP power spectra in green, and the effective noise in purple. The primordial MAP spectrum is biased low both on large scales (ISW contamination) and small scales (noise contamination). The ISW MAP spectrum is significantly biased low on all scales due to the same Wiener filter suppression. 

\begin{figure}[!htbp]
\includegraphics[width=0.9\linewidth]{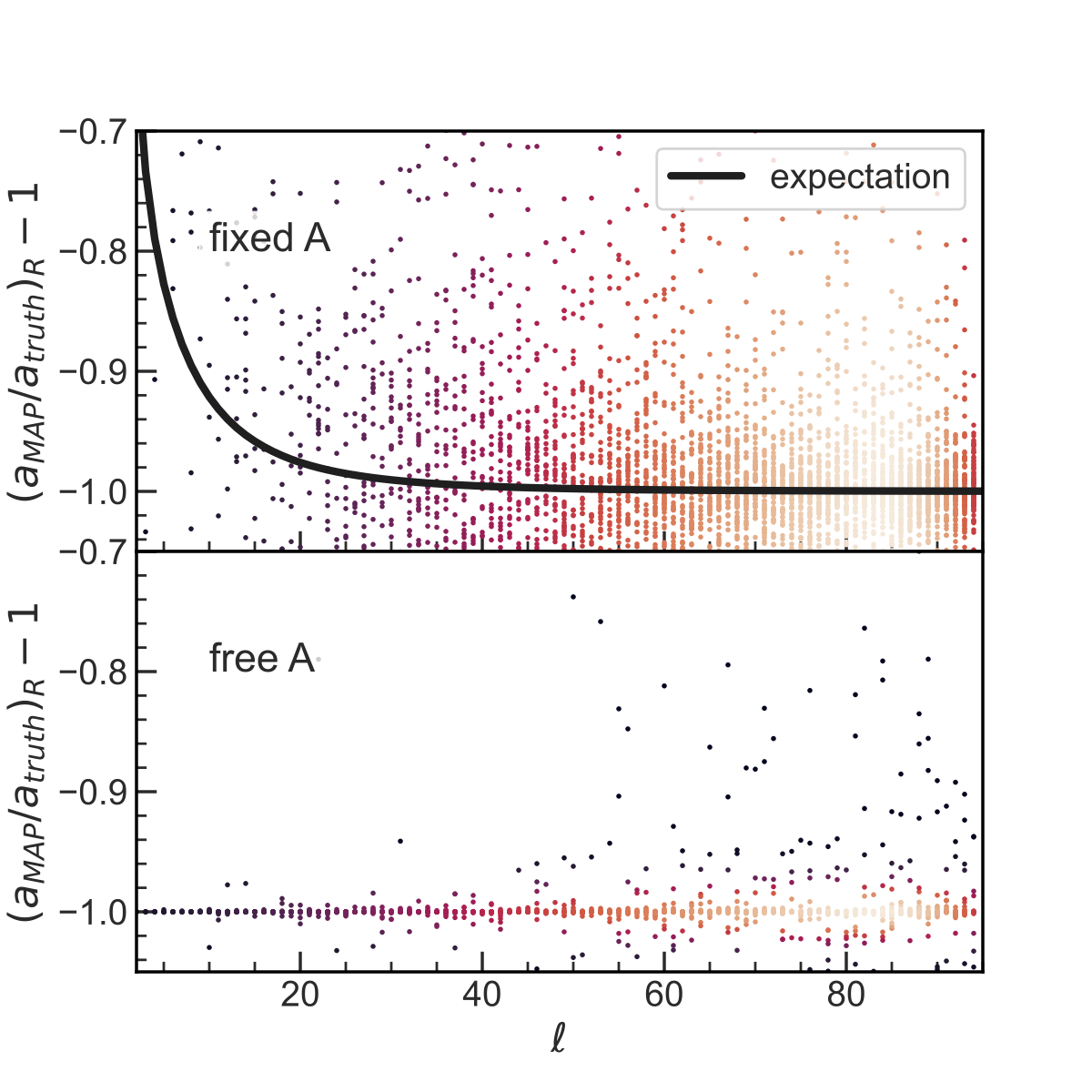} 
\caption{The scatter plot of the ratio between the MAP and the truth $a\lm$'s real components. The upper panel shows the case with fixed amplitude and the bottom shows the case with free amplitude. In the case of the fixed amplitude, the mean of the ratios is again the Wiener filter factor (black). In the case of free amplitude, the MAP solution has a null spectrum. Thus the ratios scatter around 0 for all scales.}
\label{fig:CMBISW_scatter}
\end{figure}

On the field level, we again expect (and indeed observe) no additive bias but a multiplicative bias on the $\langle \sMAP\lm/s\lm \rangle$ proportional to the Wiener filter factor (for both the real and the imaginary components). The case for the ISW field is particularly egregious, as shown in the top panel of Fig.~\ref{fig:CMBISW_scatter} (note that the $y$-axis does not even contain the unbiased case). 

\begin{figure*}
\begin{tabular}{c}
    \includegraphics[width=1\hsize]{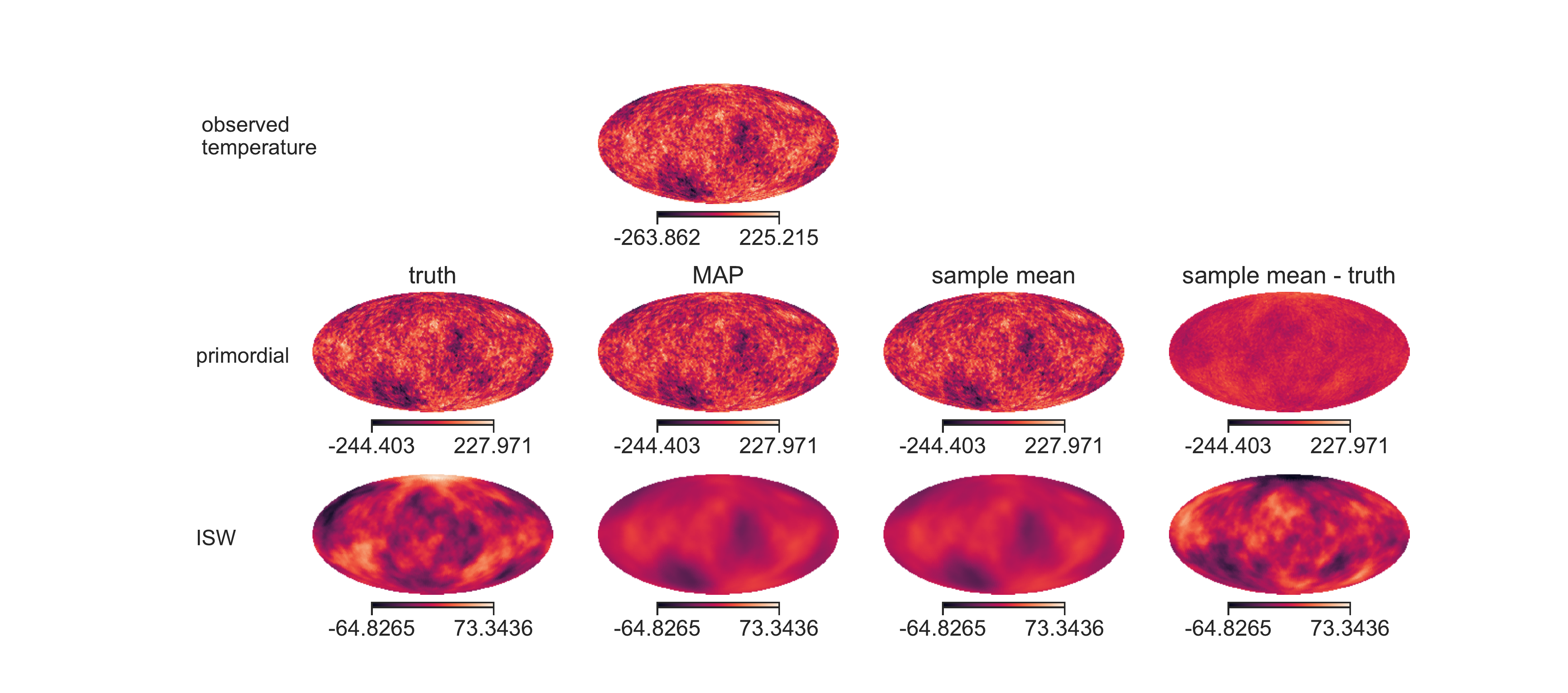}  \\
    \includegraphics[width=1\hsize]{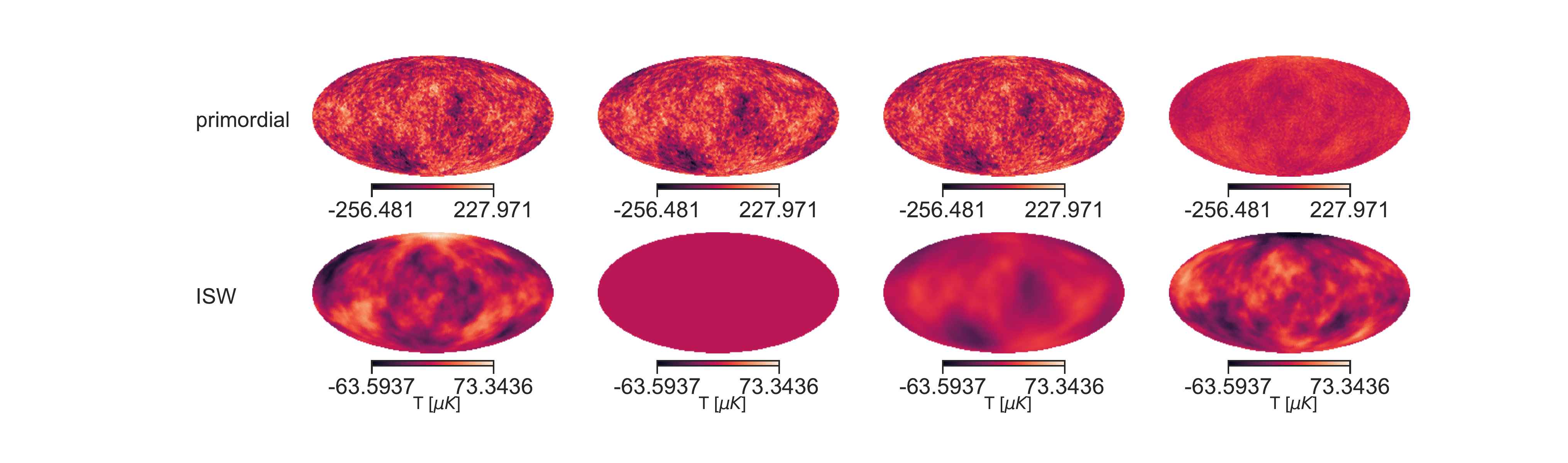} \\
\end{tabular}
\caption{Comparison between the observed data (row 1), truth maps, MAP maps, and the mean sampled maps. Rows 2-3 show the model with fixed amplitude, while rows 4-5 show the model of free amplitude. The color bar is shared across each row. Note that in the case of fixed amplitude, the mean sampled map is equal to the MAP map (second and third columns in the second and third rows), while when the amplitude is varied, the two diverge (same columns in the fourth and last row, but most obvious in the ISW maps in the last row) since the posterior marginalized over the amplitude is no longer a simple Gaussian in the signals.}
\label{fig:CMBISW_map}
\end{figure*}

Now we turn to the reconstructed real space maps (rows 1-3 of Fig.~\ref{fig:CMBISW_map}) which shed more light onto the degeneracy between the reconstructed $\sP$ and $\sI$. Qualitatively, the primordial MAP map captures most features of the true signal, although the small-scale structures are suppressed due to Wiener filtering. However, the MAP estimator completely fails in the ISW reconstruction. In fact, the $\sIMAP$ map looks like a low-pass filtered $\sP$ map. 

The physical explanation is that, when we observe a large-scale hot spot in the sky, it is impossible to confidently associate it with either the primordial CMB or the ISW effect since they both have high amplitudes at low $l$'s. However, when we optimize the posterior function with respect to $\sI$, the algorithm neglects the $\sP$ (Eq.~\ref{eqn:MAP_s_noamp_CMBISW_I}). Thus, the algorithm inclines to increase the amplitude of the $\sI$ map wherever we observe a large-scale hot spot in $\dT$, even though it is most likely due to $\sP$ since it has a greater power spectrum. As a result, the large-scale hot spots of the reconstructed $\sI$ are heavily correlated with $\sP$, even though they are spatially independent in theory. An analogous bias can be said for the reconstructed primordial map, i.e., the reconstructed primordial map is biased high where there is an ISW hot spot (although it is slightly more difficult to discern in the figure). In short, as we attempt to reconstruct two maps from a single observation using the MAP estimator, the degeneracy introduces significant bias on the field level that correlates with the two reconstructed maps. 

Now, we apply the sampler, as defined in \S\ref{sec:Methodology}, with $\sP, \sI$ as the free parameters. The two-point result is shown in the top panels of Fig.~\ref{fig:CMBISW_cl} in orange, where we observe that the spectra samples scatter around the truth unbiased, and the spectrum of the mean sampled map is equivalent to the MAP spectrum. The sampling result on the field level is shown in rows 1-3 of Fig.~\ref{fig:CMBISW_map}. Here we confirm that the mean sampled map is indeed the field-level MAP solution, and thus suffers from the same bias and degeneracy. Therefore, although the sampling approach solves the multiplicative bias on the two-point level, it is placing the right amount of power in the wrong place at the map level. 
\begin{figure}[H]
    \includegraphics[width=0.9\linewidth]{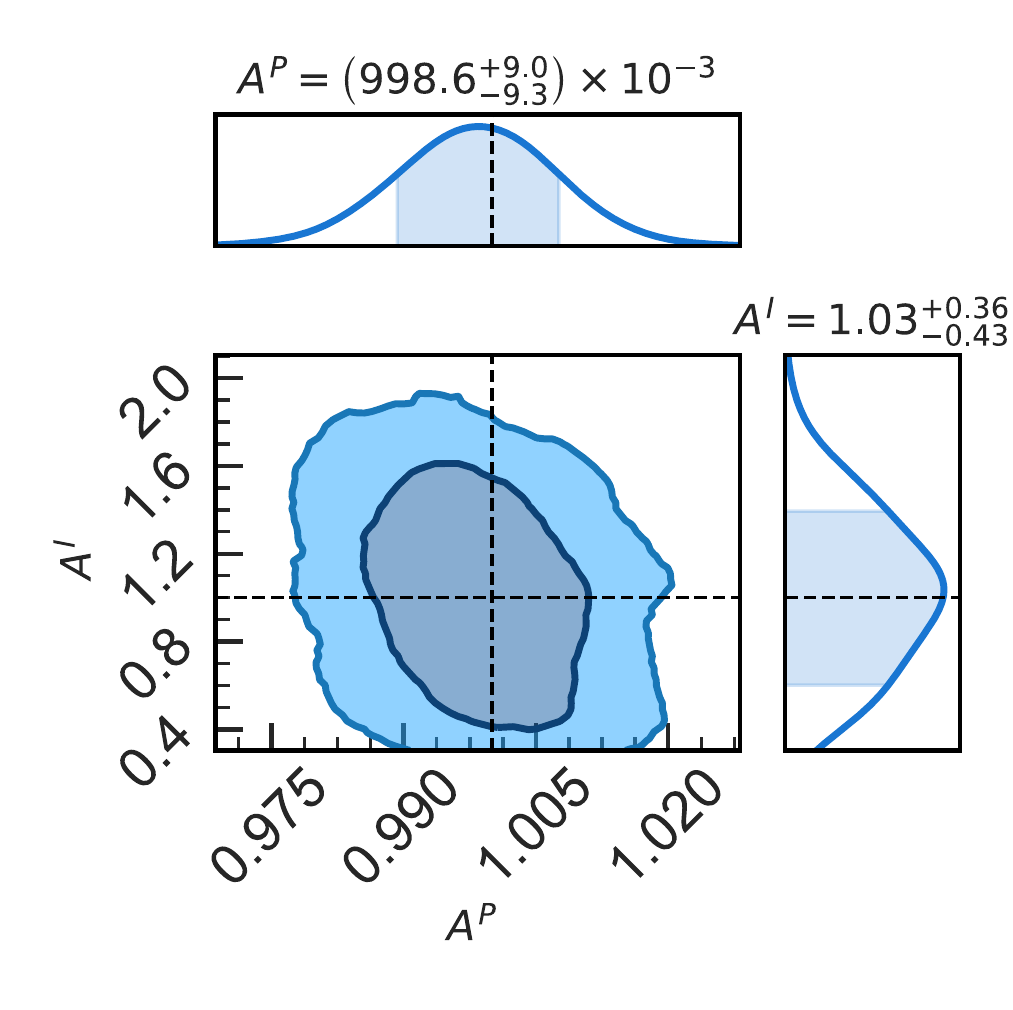} 
    \caption{The posterior distribution of the primordial and ISW amplitudes with parameter bounds. The data strongly constrain the sum of the two amplitudes (a fact not obvious here because the $x$-range is so much smaller than the $y$-range), but the sampling technique manages to correctly infer the values of both.}
    \label{fig:CMBISW_parameter}
\end{figure}

\subsubsection{Varying cosmological parameters}
We now vary the two amplitude parameters ($\AP$ and $\AI$) in Eqn.~\ref{eqn:CMBISWmodel_post} and attempt to reconstruct them together with the map pixels. The MAP solutions for $\sPMAP$ and $\sIMAP$ are analagous to Eqs. (26) and (27), but with the signal power spectra multiplied by their MAP amplitudes squared, which satisfy


\begin{align}
    \sum_{l} (2 l + 1) \left\{ 1- \frac{(\APMAP)^2 \mathbb{D}_l \CP_l}{Q_l^2}\right\}\\
    \sum_{l} (2 l + 1) \left\{ 1- \frac{(\AIMAP)^2 \mathbb{D}_l \CP_l}{Q_l^2}\right\}
    \label{eqn:MAP_s_amp_CMBISW_I}
\end{align}
where
\begin{equation}
    Q_l = (\APMAP)^2 \CP_l + (\AIMAP)^2 \CI_l + \CNT_l
\end{equation}

We apply this MAP estimator to the numerical experiment for the CMB-ISW model discussed above. The biased two-point results are shown in the bottom panels of Fig.~\ref{fig:CMBISW_cl} in green. In the case of the ISW reconstruction, the effective noise is so large that the slope of the posterior distribution with respect to the amplitude (Eq.~\ref{eqn:MAP_s_amp_CMBISW_I}) never achieves 0. This results in $\AI$ and hence the MAP spectrum being set to 0, which we confirm using direct numerical optimization. This effect is also shown in the bottom panel of Fig.~\ref{fig:CMBISW_scatter} where we plot the ratio between the MAP and truth pixel values in harmonic space. 

We also construct and apply an HMC-NUT sampler similar to the previous section but with the additional amplitude dependence. The sampled spectra and the spectrum of the mean map are shown in Fig.~\ref{fig:CMBISW_cl} in shaded orange and in red respectively. Similar to the CMB model, the sampled spectra are unbiased. This is also confirmed by the parameter constraint as shown in Fig.~\ref{fig:CMBISW_parameter}, where we see a 2.9 $\sigma$ detection of the ISW amplitude. Meanwhile, as the amplitudes are now free, the spectrum of the mean map is still biased, but to lesser degrees than the MAP solution. 

This is consistent with the field-level results (rows 4-5 of Fig.~\ref{fig:CMBISW_map}). Here, for the ISW tracer, the MAP map is essentially constant spatially, whereas the mean map still contains the right amount of power but has placed it all in the wrong place (as in the case of fixed amplitude). 

\section{Joint reconstruction of the CMB, ISW, and the galaxy density maps}
\label{sec:CMBISWGal}
We now present the main analysis, where we generalize the framework to jointly analyze data from CMB and wide-field galaxy surveys on the field level. The main goal is the following. Given an observed temperature map and six tomographic galaxy density maps, we want to construct 
estimates of the primordial ($\sP$), the ISW ($\sI$), and the galaxy density ($\{\sGi\}$) maps, along with 
two-point and cosmological parameter constraints, all in a consistent and computationally efficient Bayesian framework. 

From now on, we will only consider the sampling approach. The addition of galaxy maps introduces off-diagonal terms in the covariance of the posterior distribution, as in \ec{offdiag}.  As we shall see, following the algorithmic prescription in \S\ref{sec:Methodology}, we can break the degeneracy between the primordial and the ISW maps using additional maps of the galaxy density.

\subsection{Theoretical covariance}
\label{sec:CMBISWGal_Theoreticalcovariance}
Equation~\ref{eq:Cab} gives the general expression for the theoretical covariance of 2-dimensional tracer fields. In the case of CMB-ISW model, $\langle \sP \sI \rangle = 0$. Now, we introduce tomographic galaxy tracers, which correlate with the ISW effect (but not the primordial CMB) through their common dependence on the matter density field. This correlation will show up as off-diagonal terms in their covariance matrix, as indicated explicitly in Eq.~\ref{eq:offdiag}.

Let $N_i(\chi)$ be the normalized line of sight galaxy density distribution for the redshift bin $i$, then the galaxy clustering window function that goes into Eqn.~\ref{eqn:window} is given by
\begin{equation}
    \label{eqn:galwindow}
    W^{g,i}(\chi) = b_i N_i(\chi)
\end{equation}
where $b_i$ is the linear galaxy bias that connects the matter power spectrum and the galaxy number density power spectrum. Throughout this study, we will treat each $b_i$ as a scale-independent parameter with a fiducial value of $1$. 

\begin{figure}[!htbp]
\centering
\includegraphics[width=\columnwidth]{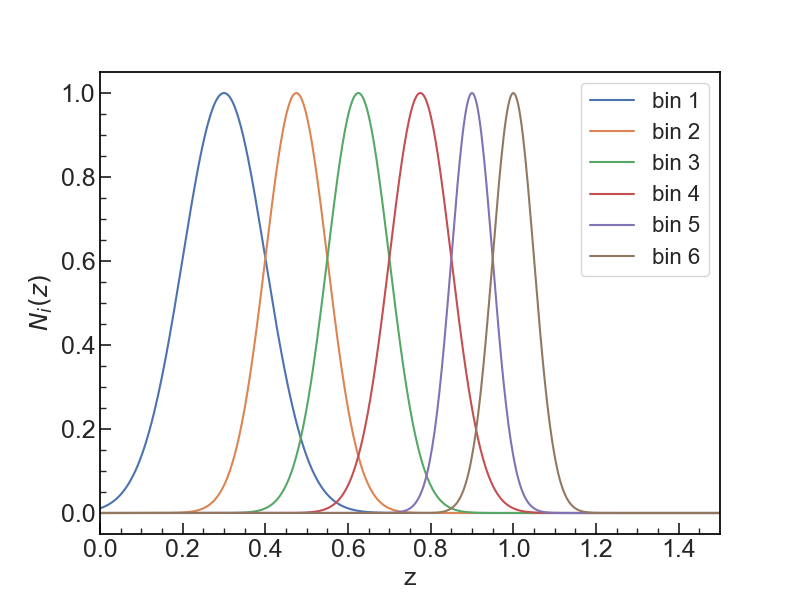} 
\caption{The galaxy number densities for each redshift bin as a function of redshift $z$. $N_i(z)$ is qualitatively modeled on the DES Year 3 \MagLim sample.}
\label{fig:nz}
\end{figure}

For wide-field photometric surveys, the galaxy redshift distribution $N_i$ varies widely between experiments and catalogs . We consider the \MagLim\ sample from DES Year3~\cite{DESY3Lens_2022,DESY3_Lensredshift_Giannini_2022} as an example; projected distributions for LSST can be found in~\cite{Boruah_2022}. The \MagLim catalog consists of 6 redshift bins spanning a redshift range of $0$ to $1.2$, calibrated using the self-organizing map methods (SOMPZ) and clustering redshifts. The catalog has been extensively tested on simulations and was used by the DES collaboration for the fiducial DES Year3 cosmology analysis \cite{DESY3_Lensredshift_Giannini_2022,Rodr_guez_Monroy_2022,DESY3Cosmology}. 
For each redshift bin, we model $N_i(z)$ using the center ($z_c$) and width ($z_w$) of the distribution, following the functional form
\begin{equation}
   \log N_i(z) \propto  -\half \left(\frac{z-z^c_i}{z^w_i}\right)^2 
\end{equation}
We also used the number density ($\Sigma_i$) of the \MagLim catalog for our simulations but assumed full-sky coverage instead of the DES footprint. We tabulate the binned $z_c$, $z_w$, and $\Sigma$ in Table~\ref{tab:zbin} and plot the normalized redshift distribution in Fig.~\ref{fig:nz}.

\begin{table}
    \centering
    \begin{tabular}{l|rrrrrr}
     & bin 1 & bin 2 & bin 3 & bin 4 & bin 5 & bin 6\\
    \hline
    $z^{c}$ & 0.30 & 0.47 & 0.62 & 0.78 & 0.90 & 1.00 \\
    $z^{w}$ & 0.10 & 0.07 & 0.07 & 0.07 & 0.05 & 0.05 \\
    $\Sigma$ [deg $^{-2}]$ & 447.29 & 319.90 & 325.48 & 435.04 & 316.74 & 298.85 \\
    \end{tabular}
    \caption{The redshift parameters.}
    \label{tab:zbin}
\end{table}

\begin{figure*}[!htbp]
\centering
\includegraphics[width=1\hsize]{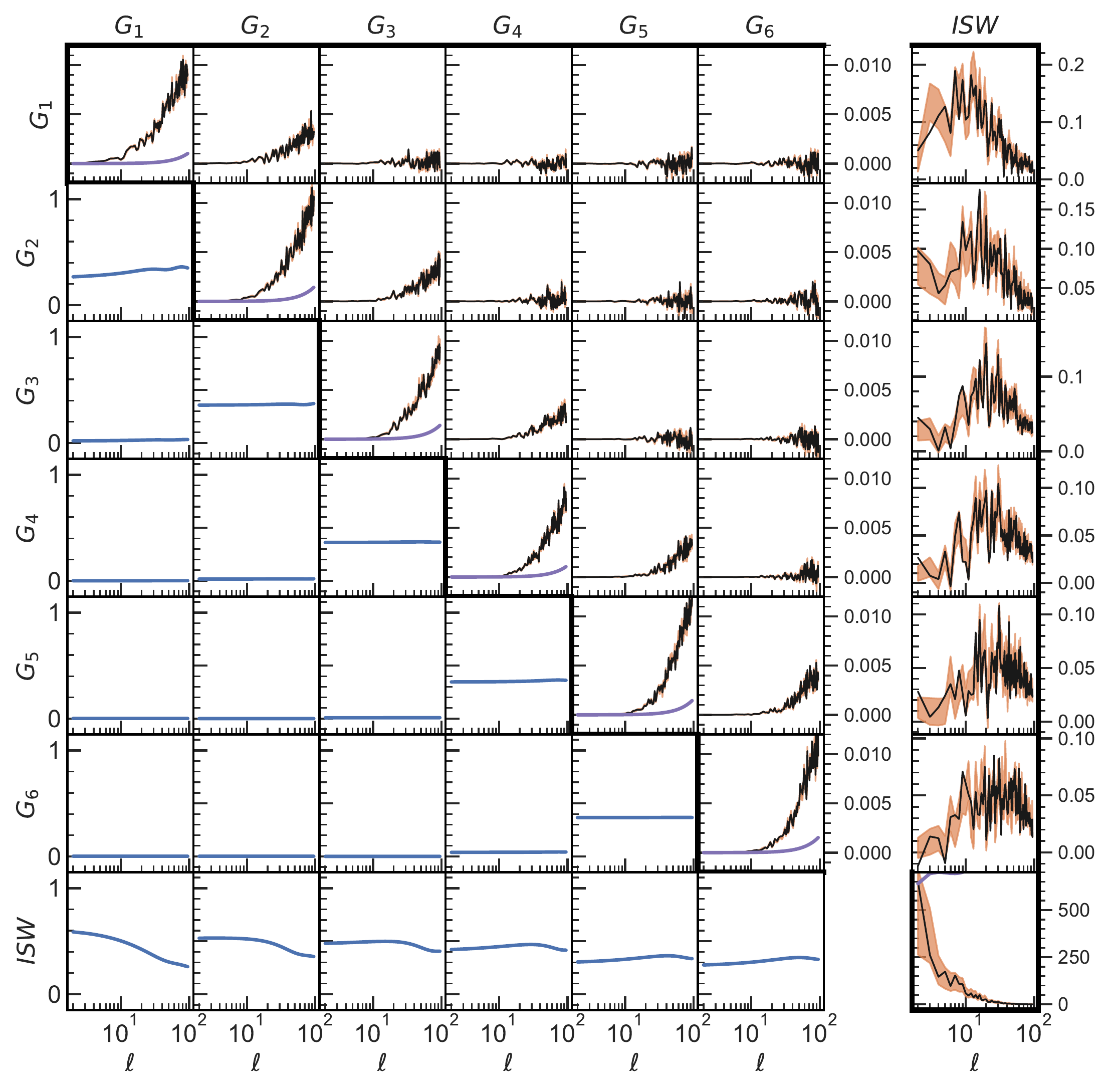}
\caption{The correlation structure between different tracers and the sampler results. The lower-left portion shows the correlation coefficients (blue) between the ISW and the 6 DES Year3 \MagLim-like galaxy tracers from different redshift bins. The primordial field is not shown here since it is independent of all late-time tracers. One crucial observation is that (lowest-left panel) the ISW effect correlates most strongly with the low redshift galaxy density maps, since only at late times did dark energy significantly drive the accelerated cosmic expansion. The upper right corner shows the distribution of power spectra sampled from the joint posterior Eq.~\ref{eqn:post_cmbiswgal} compared to the true power spectra. The input galaxy and ISW signals are shown in black while the distribution of the posterior samples is shown in orange. The diagonal subplots also contain the noise (auto)spectra. Note that the galaxy autospectrum is unitless since the galaxy redshift distribution is normalized. The ISW autospectrum has the unit of $\mu K^2$. Because of the different numeric scales, the galaxy-ISW cross-spectra and the ISW autospectrum are broken off into their own column with independent $y$-axis limits. }
\label{fig:CMBISWGal_tracerPS}
\end{figure*}

Using the window function in Eq.~\ref{eqn:galwindow}, the covariance and the correlation coefficients between the ISW effect and the galaxy density can be computed using \ec{Cab} and \ref{eqn:rhoab} (recall that primordial CMB is independent of the other tracers). The correlation coefficients are shown in the lower left corner of Fig.~\ref{fig:CMBISWGal_tracerPS} in blue. This correlation is the information we hope to leverage to break the degeneracy between the primordial and the ISW maps. The bottom-left panel demonstrates that the ISW effect is most strongly correlated with the galaxy maps on large scales at low redshift. 

\subsection{Cosmological parameters}
Let $\CP$, $\CI$, and $\CGi$ be the fiducial power spectra of the primordial, ISW, and the galaxy fields. Similar to the CMB-ISW model, we introduce an amplitude parameter for each tracer
\begin{align}
    \CP&\rightarrow  (\AP)^2 \ \CP \nonumber\\
    \CI &\rightarrow (\AI)^2 \ \CI 
    \label{eqn:galaxyamplitude}
\end{align}
Equation~\ref{eqn:galaxyamplitude} encodes a powerful stress test of \lcdm. Consider the fiducial growth function $D(z)$. If the truth cosmology deviates from \lcdm, then the actual growth function will be $A(z) D(z)$, where $A(z)$ is some redshift-dependent factor. Thus, we can interpret $\AI$ as an integral of $A(z)$ over the ISW window function. Then, any detection of $\AI \neq 1$ implies a deviation from the \lcdm model. The ISW window function is rather wide. 

The tomographic galaxy density window function is much narrower. Therefore, multiplying each binned galaxy spectrum by an amplitude factor $A^{\mathcal{G},i}$ would inform us of the consistency of \lcdm at each redshift slice. This would be, and ultimately will be, a much more strenuous test of the \lcdm model. Unfortunately, since we consider here only galaxy maps, $A^{\mathcal{G},i}$'s are entirely degenerate with the $b_i$'s in Eq.~\ref{eqn:galwindow}. So in this study, we will set $A^{\mathcal{G},i} = 1$. In future works, one could jointly analyze galaxy and shear maps to break this degeneracy and directly constrain the amplitude of the growth function in relatively narrow redshift intervals.

\subsection{Noise model and simulation}
\begin{figure*}[!htbp]
\centering
\includegraphics[width=1\hsize]{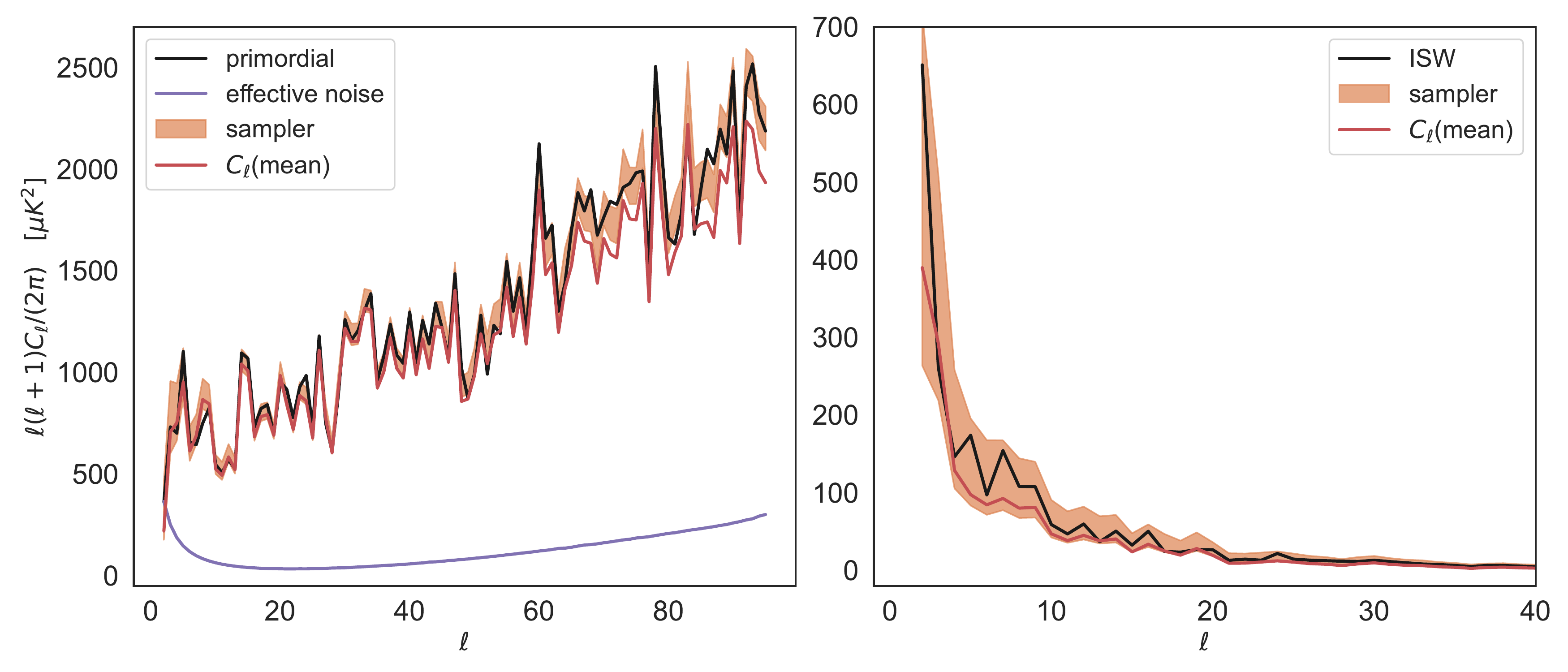} 
\caption{The recovered $\sP$ and $\sI$ spectra using the joint reconstruction technique described in \S \ref{sec:CMBISWGal}. For each case, the truth spectrum is shown in black, the $1\sigma$ credible interval of the sampled spectra is shown in shaded orange, and the spectrum of the mean map is shown in red. The effective noise of the primordial maps is $\CI+\CNT$ shown in purple. The effective noise of the CMB map ($\CP+\CNT$) is above the ISW spectra on all scales and hence not shown in the plot. Notice that both tracers are reconstructed with much higher SNR comparing to the CMB-ISW model.}
\label{fig:CMBISWGal_temperaturePS}
\end{figure*}

Since we are mostly interested in extracting the large-scale ISW signal, we again perform the simulation on a \healpix grid of $\nside = 32$. We generate the temperature and galaxy tracers using the full covariance as described in \S \ref{sec:CMBISWGal_Theoreticalcovariance}. For the observed temperature map, we inject white noise with a variance of $200~\mu K^2$. For the tomographic galaxy maps, we assume a noise spectrum of 
\begin{equation}
    \CNGi(l) = \frac{\pi^2}{180^2 \Sigma_i }
\end{equation}
where $\Sigma_i$ is the number density of bin $i$ per square degrees. The realized spectra (black) and the modeled noise spectra (purple) are shown in Fig.~\ref{fig:CMBISWGal_temperaturePS} (for the primordial and the ISW signals) and the upper right panels of Fig.~\ref{fig:CMBISWGal_tracerPS} (for the ISW and the galaxy signals). 

\subsection{Posterior modeling and sampling}
The observables are
\begin{align}
    \dT  &= \sP + \sI + \nT \\
    \dGi &= \sGi + \nGi
\end{align}
Therefore, the inference problem is specified by the posterior distribution
\begin{widetext}
\begin{align}
    \label{eqn:post_cmbiswgal}
    p(\sP, &\{\sG_{i}\}, \sI, \AP,\AI,\{b_i\} |\dT,\{\dGi\}) \propto 
     \left( \det \left[ \CNT \prod_{i=1}^6 \CNGi \right]
    \det \left[ \C_l(b_1,...,b_6,\AI,\AP)
  \right] \right)^{-\half} \notag \\
    &\times \exp{\sum\lm\frac{-|\dT\lm-\sP\lm-\sI\lm|^2}{2\CNT_l}} \quad  
      \exp{\sum_{i=1}^6 \sum\lm \frac{-|\dGi\lm-\sG\lmi|^2}{2\CNGi_l}} 
       \exp{\sum\lm \frac{-|\{\sG_{lm,1},...,\sG_{lm,6},\sI\lm,\sP\lm\}|^2}{2\C_l(b_1,...,b_6,\AI,\AP)}}
\end{align}
where  $\C_l$ is given in \ec{offdiag} and the notation $|x|^2/\C$ represents the quadratic form $x^T \C^{-1} x$
\end{widetext}

The first determinant terms encode the noise covariance, which is fixed in our example. The $\C_l$ determinant encodes the posterior's dependence on the tracer amplitudes. The first two exponential terms come from the likelihood of the temperature and the galaxy maps. The last exponential term is the Gaussian priors on the signals, including the off-diagonal covariance matrix from \ec{offdiag}. Not shown here is that all the free amplitudes have flat priors in the interval $[0.2,3]$. 

The structure of the sampler is much the same as in the CMB and the CMB-ISW models, following the prescription of  \S\ref{sec:Methodology}. Notice that since now the covariance $\C_l(b_1,...,b_6,\AI,\AP)$ is nondiagonal and extremely high dimensional, we have to employ the block diagonal Cholesky decomposition method introduced in \S\ref{sec:Methodology} to make the sampler computationally feasible. 

Despite the high dimensionality of the problem, we find that the chain equilibrates quickly (more details are given in Appendix \ref{appendix:convergence}). In general, the amplitude parameters have a much longer correlation length than latent map parameters during the sampling phase. Among the amplitude parameters, the ISW amplitude has a much longer correlation length of $10^2$ samples. Overall, the sampler is very fast. The entire analysis took less than 10 hours on a single Apple M1 CPU.

\subsection{Results}
For each iteration of the sampler after burn-in, we collect a set of maps and parameters 
\begin{equation}
    \left\{\bfs_i,{\bf A}_i\right\} = \{\sP\lmi,\sI\lmi,\{\sG_{lm,k}\}_i,\AP, \AI, \{b_k\}_i\}
\end{equation} for $k=1,...,6$ labeled by a common sample index $i$. The collection $\{\bfs_i\}$ forms the set of posterior samples which we will now analyze. We will present our findings in three parts: cosmological parameter constraints, power spectra reconstruction, and field-level reconstruction. The final results for this joint analysis are shown in Figs.~\ref{fig:CMBISWGal_tracerPS} - \ref{fig:CMBISWGal_corner}. 

\subsubsection{Cosmological parameter constraints}
\begin{figure*}
\centering
\includegraphics[width=1\hsize]{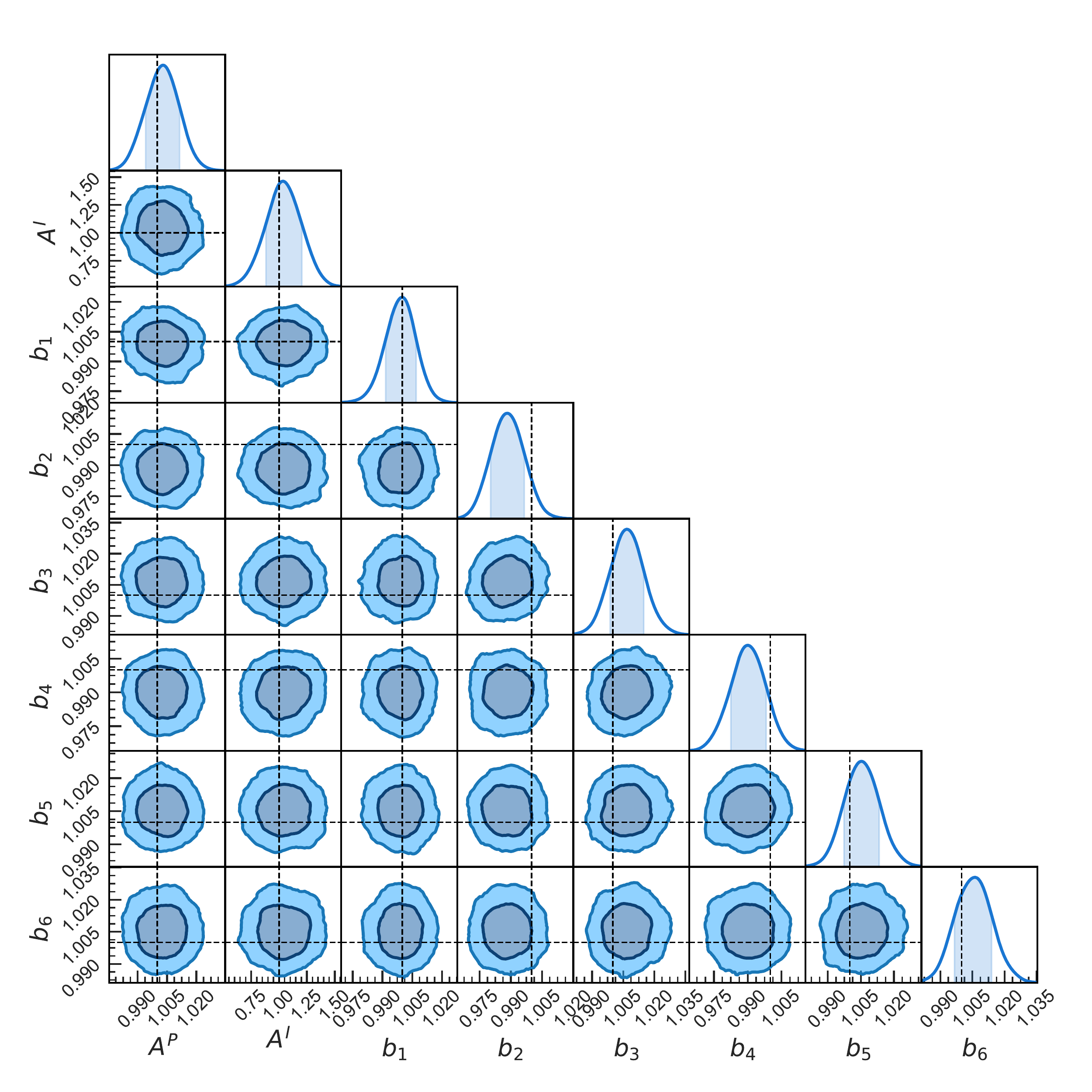} 
\caption{The posterior distribution of the amplitudes $\AP$ and $\AI$ and the $6$ galaxy biases $b_i$'s. The 2-dimensional contours label the $0.68$ and $0.95$ credible intervals. The KDE-smoothed histograms above the contours are the parameters' marginal distributions, where the shaded intervals represent the $0.16$ and $0.84$ quantiles. The truth values are indicated by dashed black lines.}
\label{fig:CMBISWGal_corner}
\end{figure*}

The constraints on the two temperature tracer amplitudes and the six galaxy biases are shown in Fig.~\ref{fig:CMBISWGal_corner} and summarized in Table~\ref{tab:model_params}. All $8$ parameters are unbiased within $2\sigma$. Consistent with our previous findings, the best-constrained parameters are the primordial amplitude and the tomographic galaxy biases which all have small effective noise. For these parameters, we achieve percent-level constraints assuming our very simple problem setups. We find that $\AI$ is also unbiased and constrained to around $15\%$ (or a $6.9\sigma$ detection), improving dramatically compared to the $\sim 40\%$ constraint (or $2.9\sigma$ detection) in the absence of galaxy data.

\begin{table}
    \centering
    \begin{tabular}{cc}
        \hline
		Parameter & Constraint \\ 
		\hline
		$\AP$ & $1.0029^{+0.0085}_{-0.0089}$ \\ 
		$\AI$ & $1.03^{+0.17}_{-0.15}$ \\ 
		$b_1$ & $1.0000^{+0.0070}_{-0.0084}$ \\ 
		$b_2$ & $0.9882^{+0.0083}_{-0.0080}$ \\ 
		$b_3$ & $1.0068^{+0.0081}_{-0.0082}$ \\ 
		$b_4$ & $0.9900^{+0.0083}_{-0.0074}$ \\ 
		$b_5$ & $1.0052^{+0.0081}_{-0.0079}$ \\ 
		$b_6$ & $1.0060^{+0.0080}_{-0.0094}$ \\ 
		\hline
    \end{tabular}
    \caption{The cosmological parameter constraints given by the CMB-ISW-galaxy model. Here $\AP$ and $\AI$ are the amplitudes of the primordial and the ISW power spectra and $b_i$ are the tomographic galaxy biases.}
    \label{tab:model_params}
\end{table}

\subsubsection{Power spectra constraints}
Besides obtaining the correct overall power spectra amplitudes, we show that the reconstructed power spectra are unbiased for all tracers for all scales. The results for the primordial CMB and the ISW effect are shown in Fig.~\ref{fig:CMBISWGal_temperaturePS}, where the sampling result is shown in shaded orange and the truth is shown in black. For ISW, we further observe that the uncertainty of the power spectrum estimation also shrinks considerably around the truth compared to the CMB-ISW model. This gain in SNR is directly attributed to the new information from the galaxy tracer fields. 

The sampling result for the ISW and the galaxy tracers (all the tracers that are correlated with each other) are shown in the upper right panels of Fig.~\ref{fig:CMBISWGal_tracerPS} in orange. Here we see that the method has captured all the auto and cross-spectra of the tracer fields. The galaxy power spectra are especially well-reconstructed, in part due to their intrinsic high SNR observations. 

\subsubsection{Field-level reconstruction}
\begin{figure*}[!]
\centering
\includegraphics[width=1\hsize]{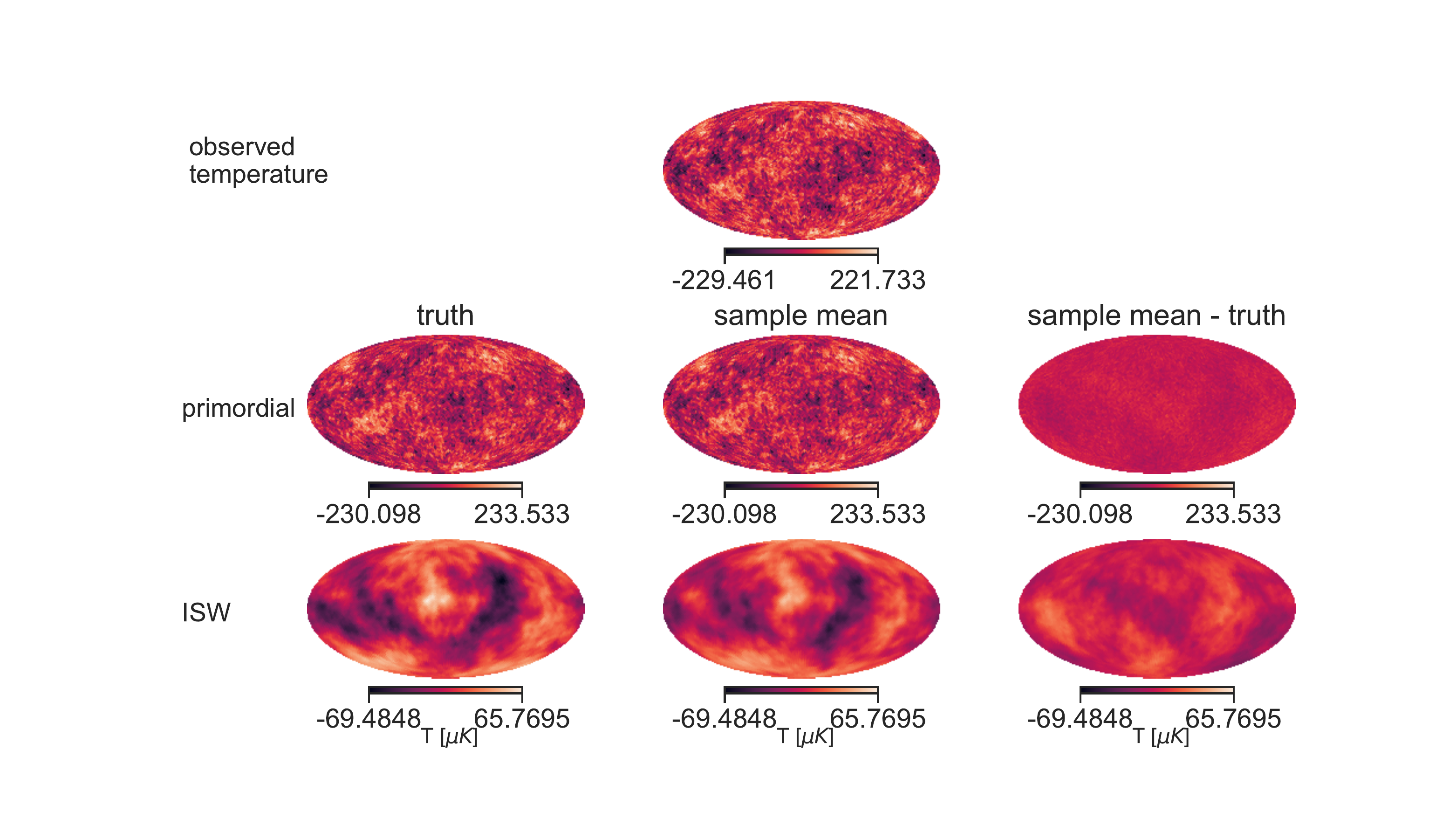}
\caption{Comparison between the observed temperature (row 1), the primordial CMB maps (row 2), and the ISW maps (row 3). For both the primordial CMB and the ISW maps, we compare the truth, the mean sampled, and the difference maps obtained from the CMB-ISW-galaxy model with free cosmological parameters.}
\label{fig:CMBISWGal_outputTmap}
\end{figure*}

The CMB-ISW-galaxy model reconstructs tracer maps at higher SNR than previous models. The result for the temperature tracers is shown in Fig.~\ref{fig:CMBISWGal_outputTmap}. Comparing to the CMB-ISW model (Fig.~\ref{fig:CMBISW_map}), we see a dramatic improvement in the reconstruction accuracy. Under the multiprobe joint reconstruction framework, the field-level information in the galaxy maps funnels into the temperature map-making process and efficiently breaks the degeneracy between the primordial and the ISW field. Most notably, although the ISW signal is by far noise dominated on all scales, the mean sampled ISW field is now actually tracing the structures of the true ISW field and decorrelated with the true primordial field. The primordial reconstruction also receives the same benefit, as the residual error of its reconstruction is visibly less correlated with the true ISW field, compared to the CMB-ISW model.

\begin{figure*}[!]
\centering
\includegraphics[width=1\hsize]{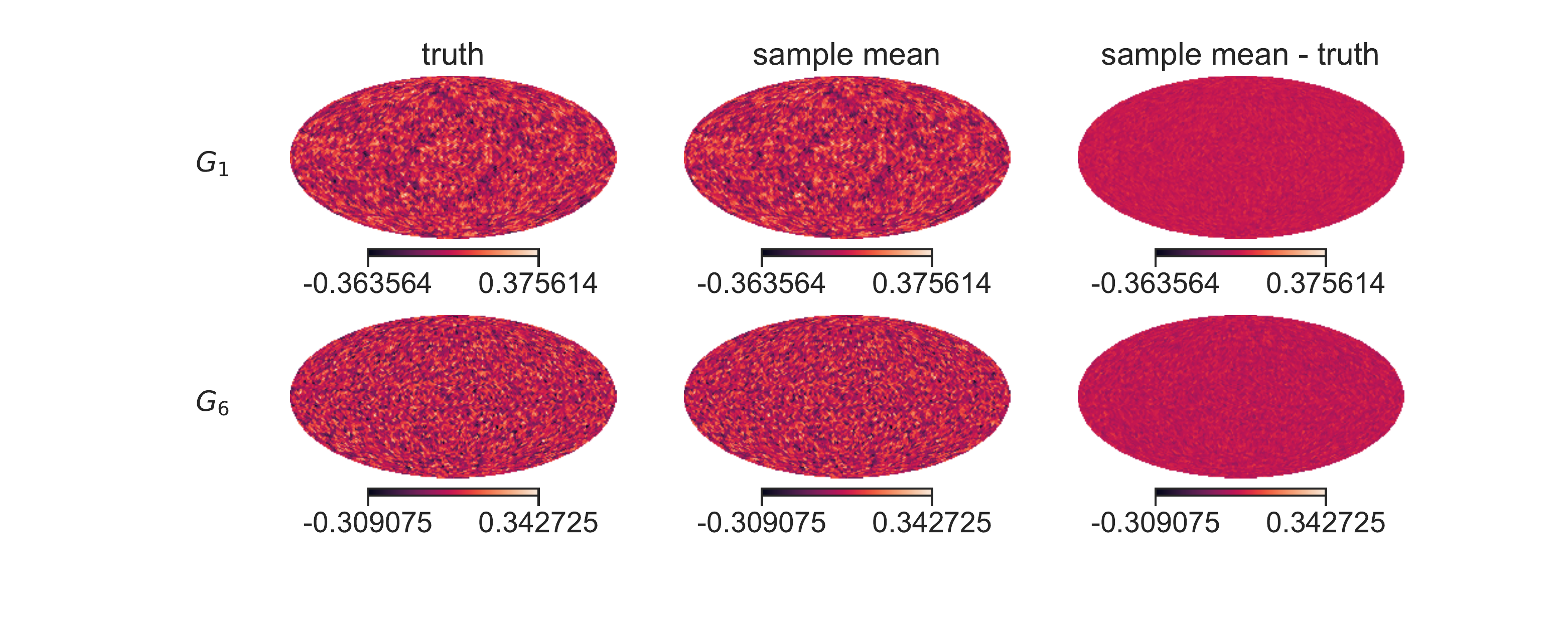}
\caption{Map-level reconstruction of the galaxy fields. Here we only show the first and the last redshift bins, the results for other bins are similar. Unit is in dimensionless overdensity. Note that although the two redshift bins have the same map resolution, the pixel scale of the lower redshift bin corresponds to smaller physical scale. Thus, the lower redshift bin has higher map-level variance, as expected. }
\label{fig:CMBISWGal_outputGmap}
\end{figure*}

The final result also includes samples of denoised tomographic galaxy maps as shown in Fig.~\ref{fig:CMBISWGal_outputGmap}. In theory, the cross-correlation with the ISW component of the temperature map can boost the SNR of the reconstructed galaxy maps as well. However, since the observed galaxy maps are already high in SNR and the degrees of freedom in the galaxy maps by far overwhelm that of the temperature tracers, their improvement is negligible. 

\section{Conclusions}		
We have implemented a general hierarchical Bayesian framework that employs HMC to sample directly from the joint posterior of the field-level multiprobe model. We did this in a pristine framework: simulated all-sky maps with simple noise properties. One goal of this is to understand the advantages and limitations of field-level analyses prior to including more realistic effects. The other is more specific: to assess how accurately surveys can measure the ISW amplitude. 

We particularly focused on comparing two approaches:  maximum {\it a posteriori} (MAP) and sampling. This enabled us to demonstrate both the well-known bias of the MAP two-point estimator~\cite{1992ApJ...398..169R} (e.g., the lower jagged curves in Fig.~\ref{fig:CMB_summary}) and the multiplicative bias of the field-level values (Fig.~\ref{fig:CMB_scatter}). These MAP biases persist as we added more complexity to the data vector and the contributing signals. The sampling approach is also biased at the field level (although, in general to a lesser extent) but is unbiased for power spectra and cosmological parameter constraints, as illustrated in Figs.~\ref{fig:CMB_A}, \ref{fig:CMBISW_parameter}, and \ref{fig:CMBISWGal_corner}. This suggests that {\it the Bayesian posterior sampler produces unbiased cosmological parameters when multiple surveys are analyzed jointly}. Given the potential biases involved in map making and then cross-correlating, this seems to us to provide an excellent justification for the use of field-level analyses moving forward.

Our results addressing the second goal can be expressed in a single number. Using only CMB data, the amplitude of the ISW signal can be extracted from the CMB only with little power, perhaps at the $2-3 \sigma$ level. However, when galaxy survey data is added, we project a $6.9 \sigma$ detection. Before exploring the limitations of this projection, it is worth emphasizing that the long-term goal is to stress-test $\Lambda$CDM, and this method provides a test at the 10-15\% level. However, this is but one of a slew of amplitudes that can be measured with upcoming data so we are optimistic about this general idea of constraining amplitudes of large-scale power spectra as a powerful way of testing the fiducial cosmological model.

Back to the limitations of our analysis: assuming full-sky slightly overstates the capability of the CMB, which is masked in the Galactic plane and overstates by at least a factor of two the coverage of, e.g., LSST. Statistically, then, one might reasonably inflate our projections by $\sqrt{2}$. However, there are a number of signals that we did {\it not} include: galaxy shapes and CMB lensing~\cite{milleaBayesianDelensingOfCMB2019,MilleaBayesianDelensingDelight2020,milleaOptimalCMBLensingFieldlevel2021}. The kernels for both of these -- especially the former -- overlap significantly with that of ISW, so we expect that including them will quite likely recover this factor of $\sqrt{2}$. By adding these observables into the analysis, we could construct tomographic (convolved with different, albeit overlapping, kernels) maps of the matter density in a consistent Bayesian framework.

However, before turning to real data, we must relax the simplifications of our simulations, so we spend the rest of this conclusion alerting ourselves and our readers to those hurdles.

\subsection{General cosmological parameters}
There is no conceptual barrier to including cosmological parameters that modify the shape of the cross-spectra (in contrast to $A$ and $b$ which modulate only their amplitudes). The main challenge is that we must specify the derivatives of the posterior distributions with respect to each of the cosmological parameters in a computationally efficient fashion (e.g., finite difference methods will be too inefficient in an inference algorithm of this scale). The process of differentiating through the Boltzmann code and the Limber approximation computation is especially difficult. We see a few ways for future projects to tackle this issue.
\begin{enumerate}
    \item Use automatic differentiation to take gradients through the cosmological dependence \cite{campagneJAXCOSMOEndtoEndDifferentiable2023}. 
    \item Use simple fitting functions (e.g., \cite{eisensteinPowerSpectraCold1999}) where the analytical derivatives are easily attainable. 
    \item Train a neural network-based cosmological emulator where the network is by definition differentiable \cite{boruahMapbasedCosmologyInference2022,nishimichiDarkQuestFast2019}. 
    \item Exclude cosmological parameters from the HMC sampling altogether. Instead, we can sample the cosmological parameters and the maps iteratively in the Gibbs sampling paradigm~\cite{Alsing_2016, wandelt_globalexactCMBGibbs2003, milleaOptimalCMBLensingFieldlevel2021}.
\end{enumerate}

\subsection{Masking, anisotropic noise, and other systematic effects}
The likelihood model we used in this study is very simple. We considered only the case of signal reconstruction on the full sky with isotropic noise and no masking. In order to adapt this algorithm for real-data analysis, we must take into account the limitation of survey geometry for both experiments, as well as foreground and point source masks. Further, the noise in real data will often be anisotropic and often a parametric function of a set of systematics spatial templates. Both the masking and the anisotropic noise models will introduce off-diagonal terms in the covariance matrix computation in Fourier space. Thus, a computationally efficient solution is to still sample the full sky, unmasked, and noiseless maps in Fourier space, transform the maps into real space, and define the likelihood there. Once in real space, we can implement different anisotropic and parametric noise models, and even attempt to constrain nuisance noise model parameters during sampling as well. 

Systematic effects such as foregrounds  and survey properties can be handled in the general framework of this field-level analysis. In particular, the {\it signal} vector can be expanded to include these. This one-step approach -- as opposed to the current treatments -- may be necessary for future surveys with increased statistical precision. One simple way to understand why this may be needed is that a sample of cosmological parameters that predicts large clustering is more likely to label an ambiguous object a galaxy (rather than a star) if it is near another galaxy. Other systematics -- such as photometric redshift uncertainty -- can be included by introducing nuisance parameters. 

\subsection{Smaller scales}
We have included only large scales here, and there is a huge advantage to doing so, in that the prior distributions of the signals is known to be Gaussian. There is a huge disadvantage to throwing out all the information available on small scales. Including small scales in the posterior requires a knowledge of the prior distribution of the signal, a distribution that is less and less Gaussian as we push to smaller scales. There are two possible approaches to this: (i) assume a simple distribution (e.g., Gaussian or log-normal~\cite{boruahMapbasedCosmologyInference2022}) and investigate the potential biases by running the pipeline on simulations and (ii) the ambitious approach of rolling the clock back and using the primordial fields as the parameters given the observed highly processed fields~(e.g., \cite{https://doi.org/10.48550/arxiv.2210.15649,https://doi.org/10.48550/arxiv.2104.12864}). 

\section*{Acknowledgements}
This work is supported by U.S. Dept. of Energy Contract No. DE-SC0019248 and by NSF Award Number 2020295. S.D. is grateful to the Aspen Center for Physics, where a workshop in Summer 2022 exposed him to some of these ideas. We are also grateful to Alan Heavens, Andrew Jaffe, Xiangchong Li, Marius Millea, Chirag Modi, Fabian Schmidt, and Ben Wandelt for useful conversations.

\bibliographystyle{mnras}
\bibliography{refs}

\appendix
\section{Distributions of individual modes of the sampled power spectra}
\label{appendix:Cl}
\begin{figure}[!htbp]
\centering
\includegraphics[width=1\hsize]{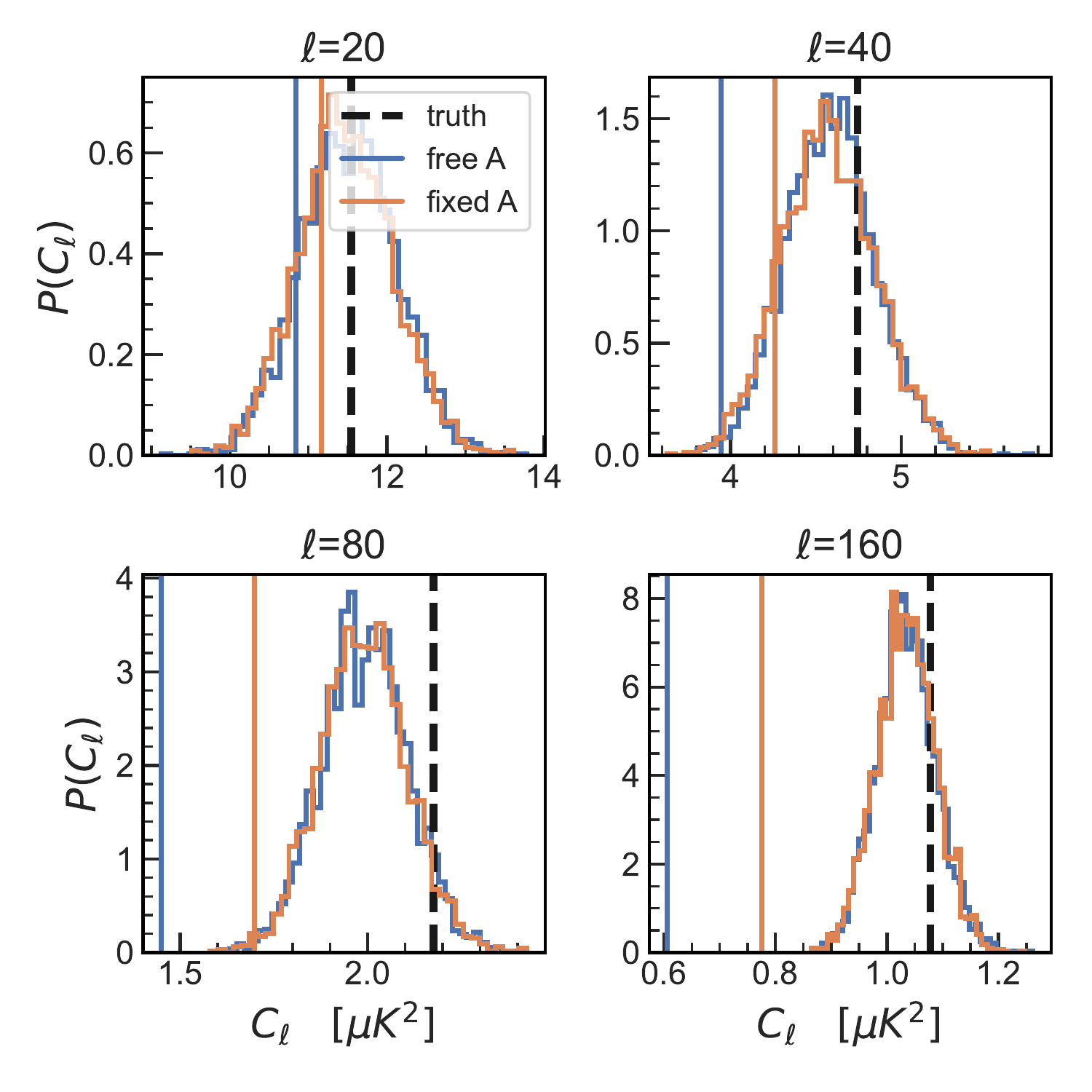} 
\caption{The marginal distribution of $\CP$ of the posterior samples for the CMB model (\S\ref{section:CMB}). The case with fixed fiducial power spectrum is shown in orange, and the case where one admits a free amplitude is shown in blue. The vertical black lines show the truth $\CP$, while the colored vertical lines show the MAP values for $\CP$.}
\label{fig:CMB_Cldist}
\end{figure}

\begin{figure}[!htbp]
\centering
\includegraphics[width=1\hsize]{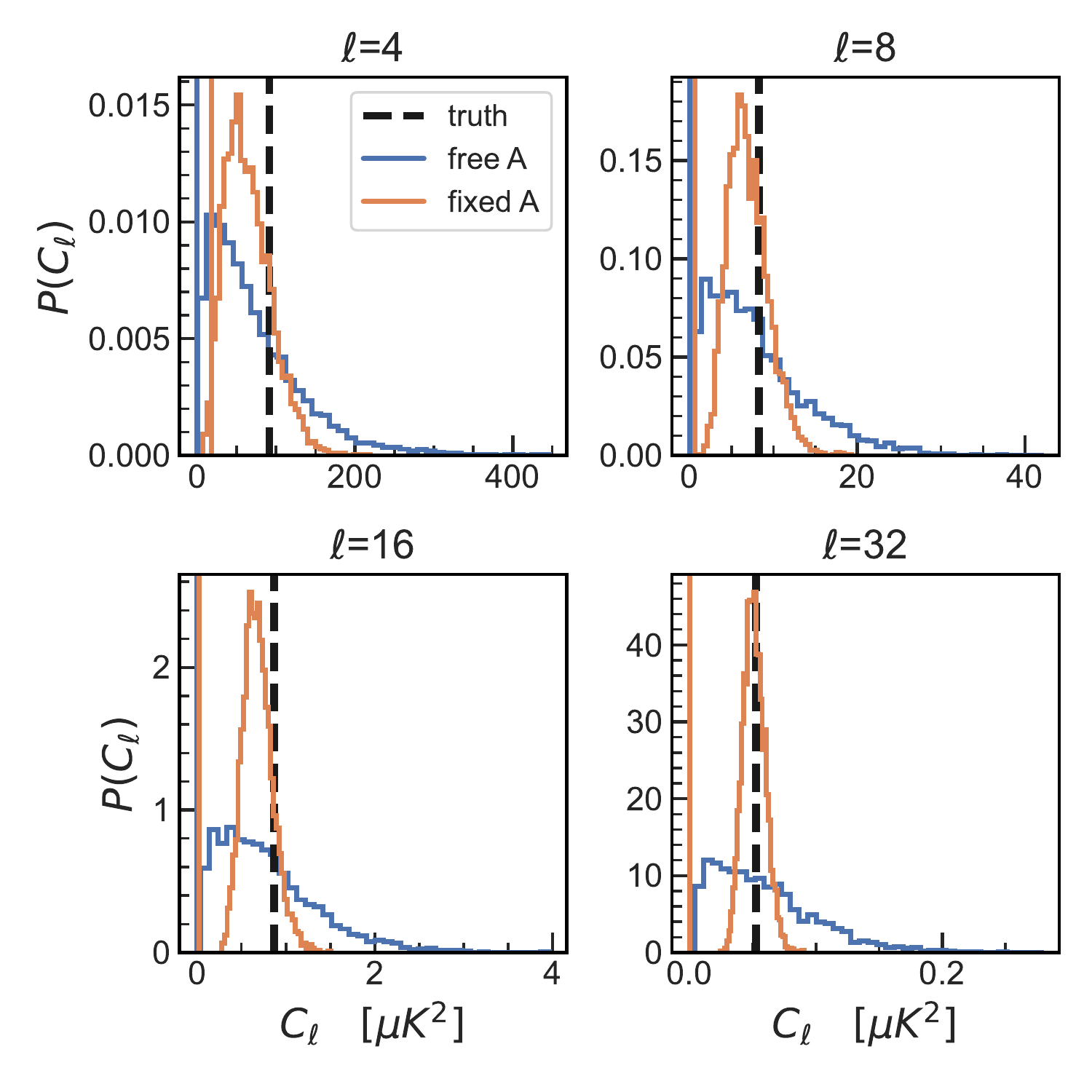} 
\caption{The marginal distribution of $\CI$ of the posterior samples for the CMB-ISW model (\S\ref{sec:CMBISW}). The case with fixed fiducial power spectrum is shown in orange, and the case where one admits a free amplitude is shown in blue. The vertical black lines show the truth $\CI$, while the colored vertical lines show the MAP values for $\CI$.}
\label{fig:CMBISW_Cldist}
\end{figure}

\begin{figure}[!htbp]
\centering
\includegraphics[width=1\hsize]{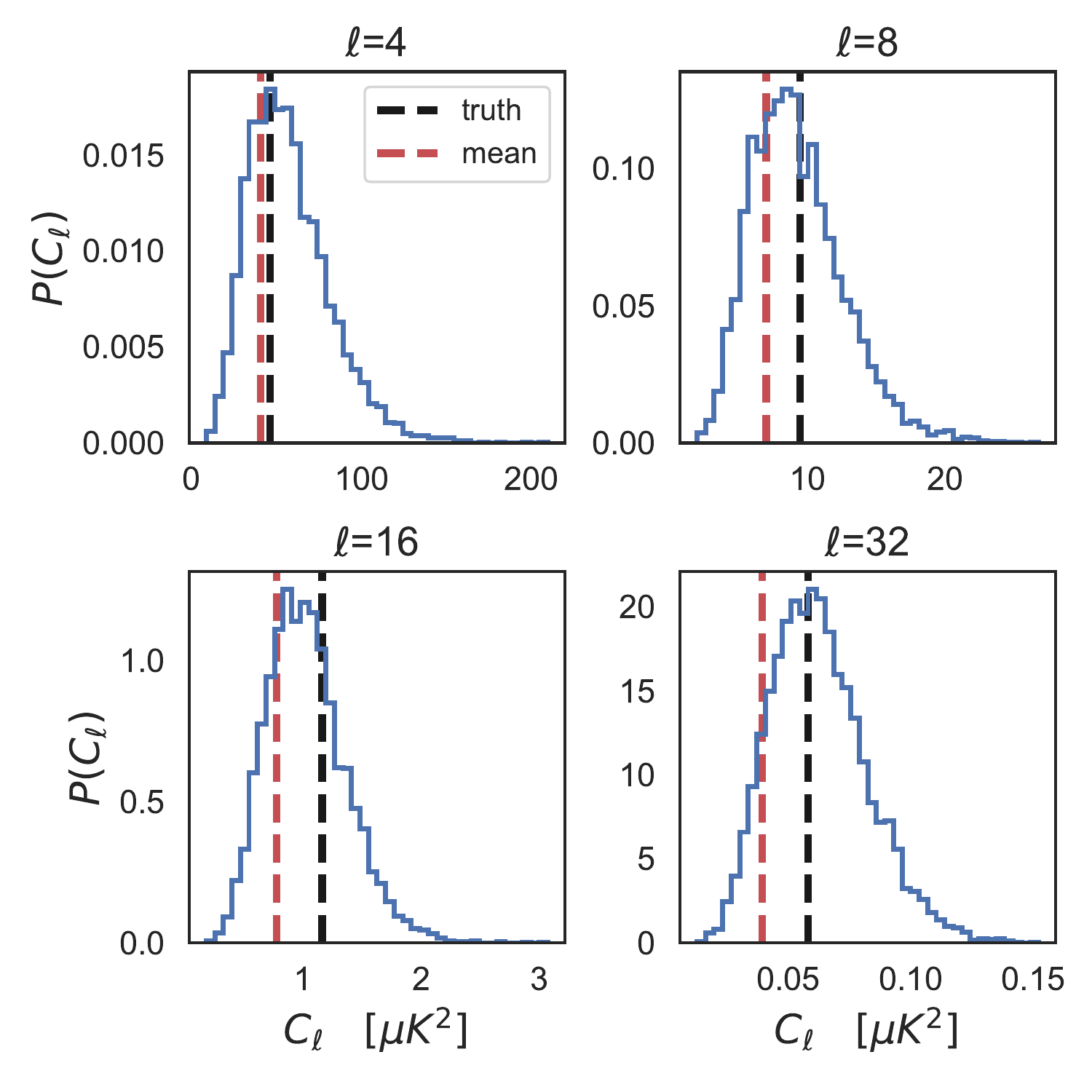}
\caption{The blue lines show the marginal distribution of $\CI$ of the posterior samples for the CMB-ISW-galaxy model (\S\ref{sec:CMBISWGal}). The black and the red dashed lines mark the truth value and the mean of the sample distribution respectively.}
\label{fig:CMBISWGal_Cldist}
\end{figure}

Here we present the marginal posterior distribution of $\C$ for each individual $l$ mode for the CMB model ($\CP$, Fig.~\ref{fig:CMB_Cldist}), CMB-ISW model ($\CI$, Fig.~\ref{fig:CMBISW_Cldist}), and the CMB-ISW-galaxy model ($\CI$, Fig.~\ref{fig:CMBISWGal_Cldist}). 

Figures~\ref{fig:CMB_Cldist} and \ref{fig:CMBISW_Cldist} demonstrate that the MAP power spectra amplitudes (vertical blue and orange lines) have greater bias 1) at higher $l$-modes where the noise power is larger 2) when the power spectra amplitudes are set free. Figure~\ref{fig:CMBISW_Cldist} and \ref{fig:CMBISWGal_Cldist} show that, in the case of free amplitude parameters, the quality of $\CI$ reconstruction (the width of its marginal distribution) improves dramatically when one introduces galaxy information.

\section{MAP and sampler estimators}
\label{appendix:proof}
Our algorithm generates samples of the posterior. In the text we made two claims about these samples (for the case of fixed cosmological parameter): (i) the mean (two-point and $n$-point) power spectrum of the samples is unbiased and (ii) in the single component case, the mean map is equal to the MAP map. To demonstrate this, 
we let $\{s^\alpha_{lm,i}\}$ be the set of maps that we have collected during sampling, where $i$ denotes the sample index and $\alpha$ denotes the type of cosmological probe. Let $N$ be the total number of samples.

\subsection{Power spectrum}
We first show that the mean sampled $n$-point power spectrum estimator is unbiased. Let us consider the pixel variance, which is the simplest two-point statistic. For a fixed data set $D$, the mean pixel variance across all the posterior samples is $\text{mean}(\langle s^2 \rangle)$. Here, $s$ is the signal map vector, $\langle s^2 \rangle$ denotes the average pixel variance of $s$, and the "mean" operation is taken over all the posterior samples. Using $i$ as the index for the posterior samples,
\begin{align}
    &\text{mean}(\langle s^2 \rangle)\\
    &= \lim_{{N\to\infty}} \sum_{i=1}^N \frac{\langle s_i^2 \rangle}{N}\\
    &= \int ds ~ |s^2| ~ p(s|D)
    \label{eqn:proof_intermediate}
\end{align}
where the short hand $|s^2|$ means $\sum_{\alpha=1,...,N_{pix}} s_\alpha^2/N_{pix}$ with $\alpha$ being the pixel index. Equation~\ref{eqn:proof_intermediate} in general depends on the data $D$. Practically, this means that $\text{mean}(\langle s^2 \rangle)$ is biased only in the sense that the observed data has intrinsic randomness. If we proceed to integrate over the intrinsic variance in the data, we have
\begin{align}
    &\int ds \int dD ~ |s^2| ~ p(s|D) ~ p(D) \\
    &=\int ds ~ |s^2| ~ p(s) \\
    &=\langle s^2 \rangle
\end{align}
which is the true pixel variance the signal. This argument works for the covariance between two different pixels, and more generally, for the $n$-point correlation functions. 

\begin{figure}[!]
\centering
 \includegraphics[width=1\hsize]{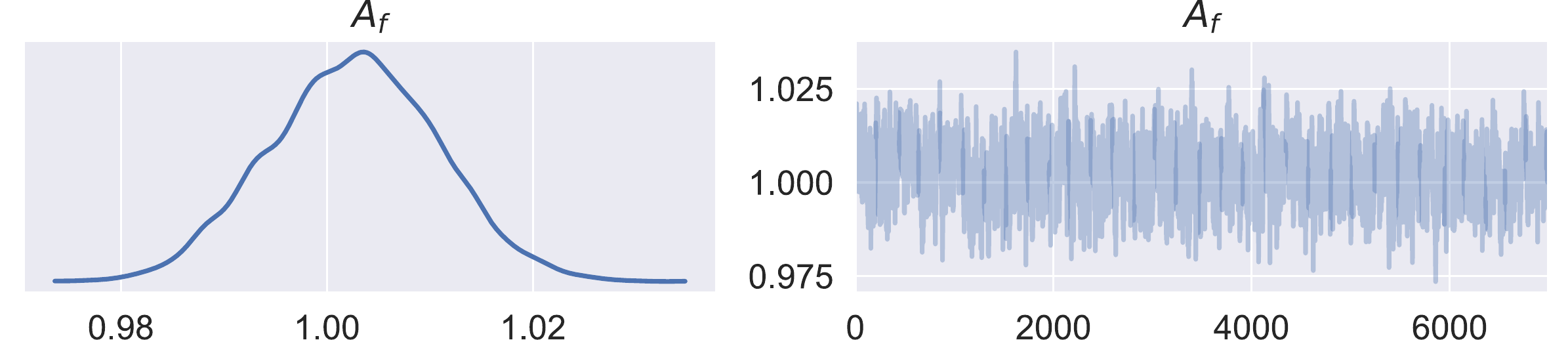}
 \includegraphics[width=1\hsize]{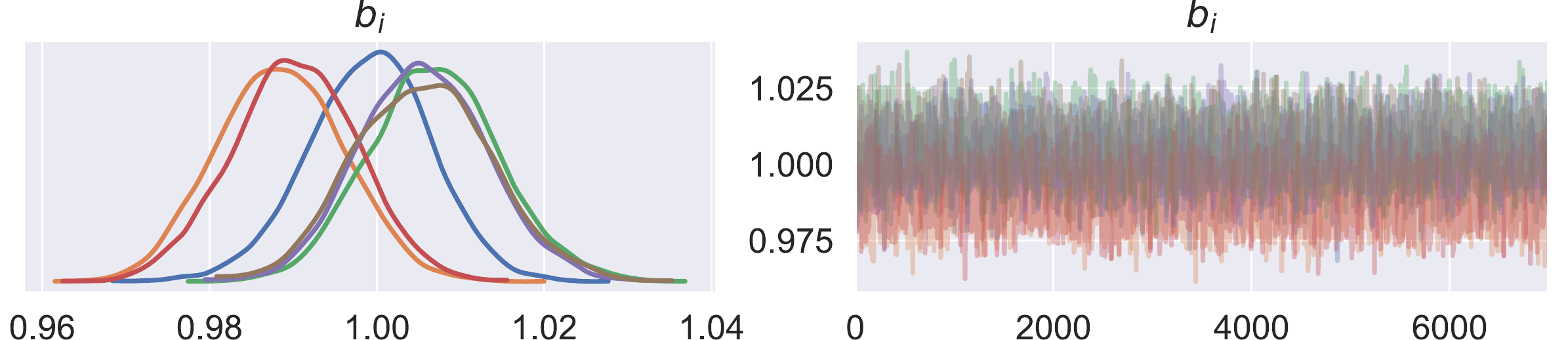}
 \includegraphics[width=1\hsize]{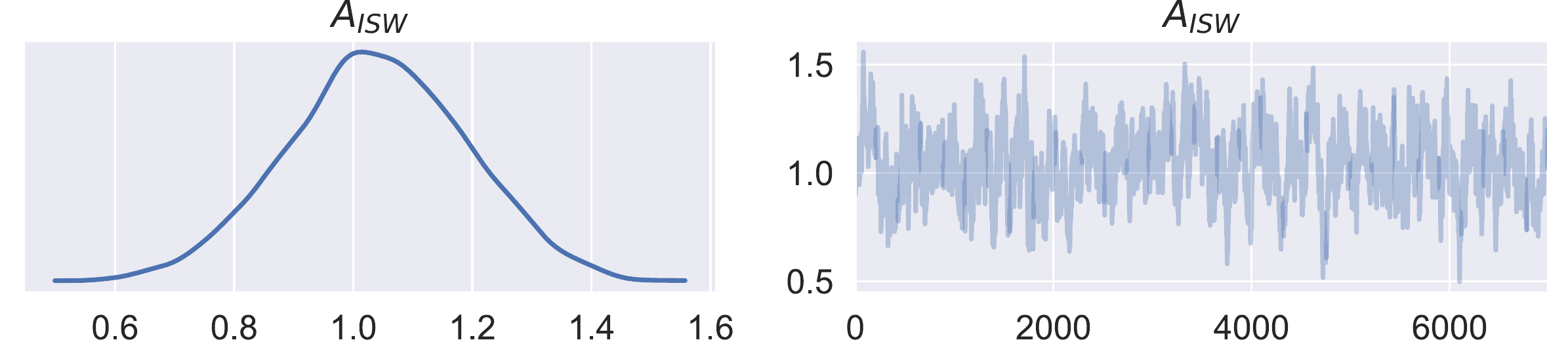}
\caption{The chain convergence of the power spectra amplitudes and galaxy biases for the CMB-ISW-galaxy model. The left panels show the posterior distribution of each parameter after KDE smoothing. The right panels show each parameter's value as a function of sample index after 3000 warmup steps. }
\label{fig:CMBISWGal_chain}
\end{figure}

\begin{figure}[!]
\centering
\includegraphics[width=1\hsize]{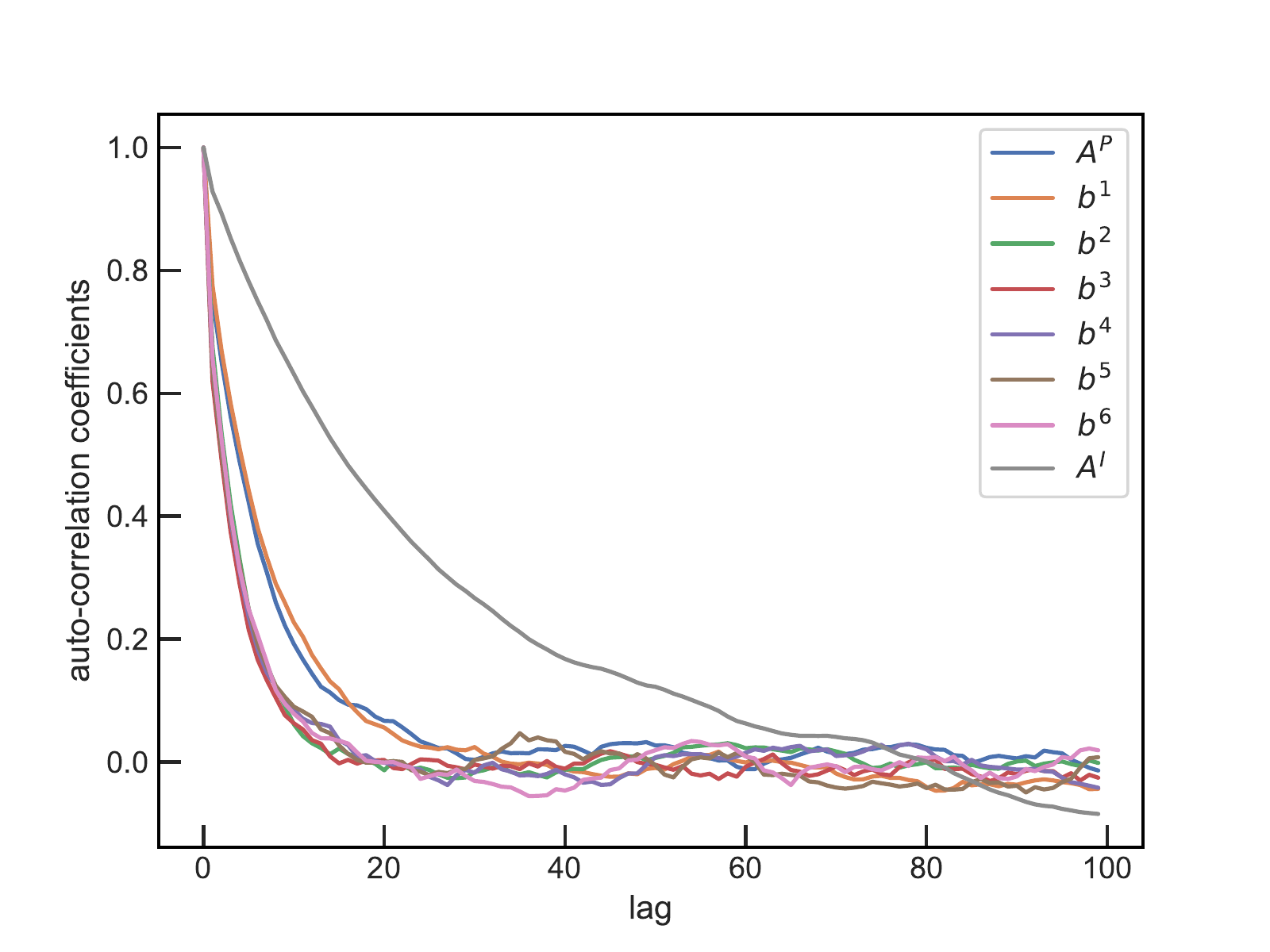}
\caption{The auto-correlation functions of the $8$ sampled cosmological parameters shown in Fig.~\ref{fig:CMBISWGal_chain}.}
\label{fig:CMBISWGal_chain_autocorr}
\end{figure}

\subsection{MAP estimator}
We claim that the mean sampled map is the field-level MAP solution, and the power spectrum of the mean map is the MAP spectrum. Similar to the argument above, we start with 
\begin{equation}
    \langle s \rangle = \int ds ~ s ~ p(s|D)
\end{equation}
where $p(s|D)$ is Gaussian. Since the mean of a Gaussian distribution is also the point of maximum probability, $\langle s \rangle$ is thus also the MAP estimator. More explicitly, by independence between different $l,m$-modes, we can suppress the $l,m$ subscript on $s$ and look at the term
\begin{align}
    & \text{mean}(s_i) \times p(D) \\
    &= \lim_{N\to\infty} \sum_{i=1}^N \frac{s_i}{N} \times p(D)\\
    &= \int ds \frac{s}{2\pi \sqrt{\C^n \C^s}} \exp \left(-\frac{(D-s)^2}{2\C^n} - \frac{s^2}{2\C^s} \right) \\
    &= \frac{\C^s D}{\C^n +\C^s} \times \frac{\exp \left(-\frac{D^2}{2(\C^n+\C^s)} \right)}{(2\pi (\C^n+\C^s))^{1/2}} 
\end{align}
which is the Wiener solution for $s$ times the probability of the data. Dividing by $p(D)$ yields
\begin{equation}
     \langle s \rangle = \frac{\C^s D}{\C^n +\C^s}
\end{equation}
as desired. It follows that the power spectrum of the mean map is the Wiener filter spectrum.

Interestingly, notice that the above arguments rests on the symmetry of the Gaussian distribution. If we introduce an additional amplitude parameter $A$, the distribution over $A$ is no longer Gaussian, and so the mean map is no longer the MAP map solution (as discussed at length in the main text).

\section{Chain convergence}
\label{appendix:convergence}

Here we present the chain convergence information of the CMB-ISW-galaxy model, focusing on the 8 cosmological parameters. The sampled parameter values as a function of sample index is shown in Fig.~\ref{fig:CMBISWGal_chain}, and the corresponding autocorrelation functions are shown in Fig.~\ref{fig:CMBISWGal_chain_autocorr}. In general, we observe that the map parameters (not shown here) have much shorter correlation lengths than the cosmological parameters. Among the cosmological parameters, $\AP$ and $b_i$ have much shorter correlation lengths compared to $\AI$.

\end{document}